\documentclass[journal]{IEEEtran}
\usepackage{amsmath}
\usepackage{graphicx}

\usepackage[dvips]{color}
\usepackage{algorithm}
\usepackage{algorithmicx}
\usepackage{algpseudocode}
\usepackage{color}
\usepackage{cite}

\newcommand{\minorchange}[1]{#1}

\newtheorem{property}{Property}[section]

\newtheorem{definition}{Definition}[section]

\newcommand{\MYTAB}{\hspace*{0.4cm}}
\newcommand{\SP}{\State}
\newcommand{\ST}{\State\MYTAB}
\newcommand{\STT}{\State\MYTAB\MYTAB}
\newcommand{\STTT}{\State\MYTAB\MYTAB\MYTAB}

\newcommand{\CODE}{\rmfamily}

\newcommand{\subbegin}[2]{
{\footnotesize\itshape #1}
\begin{algorithmic}
\footnotesize
\itshape
}

\newcommand{\subend}{
\end{algorithmic}
}

\newcommand{\msgbegin}[2]{
\begin{algorithm}
\floatname{algorithm}{Message}
\caption{{\CODE \mbox{#1}}}
\label{#2}
\begin{algorithmic}
}

\renewcommand{\algorithmiccomment}[1]{//#1}
\renewcommand{\algorithmicend}{\textcolor{blue}{\textbf{end}}}
\renewcommand{\algorithmicif}{\textcolor{blue}{\textbf{if}}}
\renewcommand{\algorithmicthen}{\textcolor{blue}{\textbf{then}}}
\renewcommand{\algorithmicelse}{\textcolor{blue}{\textbf{else}}}
\renewcommand{\algorithmicfor}{\textcolor{blue}{\textbf{for}}}
\renewcommand{\algorithmicforall}{\textcolor{blue}{\textbf{for all}}}
\renewcommand{\algorithmicdo}{\textcolor{blue}{\textbf{do}}}
\renewcommand{\algorithmicwhile}{\textcolor{blue}{\textbf{while}}}
\renewcommand{\algorithmicloop}{\textcolor{blue}{\textbf{loop}}}
\renewcommand{\algorithmicrepeat}{\textcolor{blue}{\textbf{repeat}}}
\renewcommand{\algorithmicuntil}{\textcolor{blue}{\textbf{until}}}
\renewcommand{\algorithmicprocedure}{\textcolor{blue}{\textbf{procedure}}}

\title{An Analytical Study of a Structured Overlay in the \minorchange{Presence} of Dynamic Membership 
\thanks{This work is funded by the European 6th FP EVERGROW project.
 \copyright  IEEE. Personal use of this material is permitted. However,
  permission to reprint/republish this material for advertising or
  promotional purposes or for creating new collective works for
  resale or redistribution to servers or lists, or to reuse any
  copyrighted component of this work in other works, must be
  obtained from the IEEE.}
}
\author{
          Supriya Krishnamurthy$^{1,\minorchange{3}}$, Sameh El-Ansary$^{1}$, Erik Aurell$^{1,2}$ and Seif Haridi$^{1,3}$\\
          $^1$ Swedish Institute of Computer Science (SICS), Sweden\\
          $^2$ Department of Physics, KTH-Royal Institute of Technology, Sweden\\
					$^3$ IMIT, KTH-Royal Institute of Technology, Sweden\\
        	\{supriya,sameh,eaurell,seif\}@sics.se \\     	
}

\begin{document}
\maketitle

\begin{abstract}
In this paper we present an analytical study of dynamic membership (aka churn) in structured peer-to-peer networks. We use a fluid model approach to describe steady-state or transient phenomena, and apply it to the Chord system.
For any rate of churn
and stabilization rates, and any system size, we accurately account for the functional form of the probability of network disconnection \minorchange{as well as} the fraction of failed or 
incorrect successor and finger pointers. \minorchange{We} show how we can use these quantities to predict both the performance and consistency of lookups under churn.  
All theoretical predictions match simulation results. 
The analysis includes both features that are 
generic to structured overlays deploying a ring as well
as Chord-specific details, and opens the door to a systematic
comparative analysis of, at least, ring-based
structured overlay systems under churn.
\end{abstract}

\section{Introduction}
\PARstart{A}{n} intrinsic property of Peer-to-Peer systems is the 
process of never-ceasing dynamic membership.
Structured Peer-to-Peer Networks (aka Distributed Hash Tables (DHTs)) 
have the underlying principle
of arranging nodes in an overlay graph of known topology and diameter. 
This knowledge results in the provision
of performance guarantees. However, dynamic membership continuously 
``corrupts/churns'' the overlay graph and
every DHT strives to provide a technique to ``correct/maintain'' the graph
in the face of this perturbation.

Both theoretical and empirical studies have been 
conducted to analyze the performance of DHTs undergoing
``churn'' and simultaneously performing ``maintenance''. Liben-Nowell 
 \textit{et al.} \cite{nowell02analysis} prove a lower bound on the 
maintenance rate required for a network to remain connected 
in the face of a given dynamic membership rate. 
Aspnes  \textit{et al.} \cite{aspnes02FaultTolerant} give upper and lower 
bounds on the number of messages needed to locate  a node/data item 
in a DHT in the presence of node or link failures. 
The value of such theoretical studies is that they provide 
insights neutral to the details of any particular DHT.  
Empirical studies have also been conducted to
complement these theoretical studies by  showing how 
within the asymptotic bounds, the performance of a DHT may 
vary substantially depending on 
different DHT designs and implementation decisions. 
Examples include the work of: 
\minorchange{Li \textit{et al.}} \cite{dhtcomparison:infocom05}, 
Rhea \minorchange{\textit{et al.}} \cite{rhea04handling}, 
and Rowstron \minorchange{\textit{et al.}} \cite{rowstron04depend}.


In this paper, we present a fluid model of
Chord \cite{chord:ton}, a specific DHT, under churn.
Fluid models have been used to model data communication systems at
least since the early '80ies~\cite{AnickMitraSondhi},
and in some sense since the work of Erlang~\cite{Erlang}.
More recently, in the context of P2P systems, it has been used to
model the performance 
of BitTorrent~\cite{qiu} and the Squirrel 
caching system~\cite{clevenot}.
This technique has much in common with macroscopic
and mesoscopic descriptions of physical and chemical phenomena
(from where the term fluid has obviously been borrowed), and 
carries the same advantages of conciseness
and computability relative to an underlying more
exact description. Our analysis is directly 
based on the master equation approach of physical
kinetics, see \textit{e.g.} the text book \cite{vanKampen}, which 
provides a scheme for taking the various dynamical
processes involved systematically into account.

The fluid model  requires the notion of
a {\it state} of the system. This is just a listing of the
quantities one would need to know for a description of the
system at a given level of detail. For Chord, we use {\it grosso modo}
a level of description which requires keeping track of
how many nodes there are in the system and
what the state (whether correct, incorrect or failed) of each of 
the pointers  of those nodes is. This information 
is not enough to draw a unique
graph of network-connections because, for example,
if we know that a given node has an 'incorrect' successor pointer, this 
still does not tell us which node it is pointing to. However,
as we will see,  beginning at this level of description 
is sufficient to keep track of most of the details of the Chord protocols.
Having defined a state, the fluid model is simply a set of equations
for the evolution of the probability of finding the system in this state,
given the details of the dynamics. 
The master equation approach is useful for 
keeping track of the contribution 
of all the events which can bring about changes in the 
probability in a micro-instant of time {\it i.e.},   
evaluating all the terms in the dynamics leading to a 
gain or loss of this probability. 

Using this formalism we investigate a probabilistic model in which
peers arrive independently, distributed as a Poisson process, 
and life-times are exponentially distributed.
While this setup is not necessary fully realistic (more realistic models can also be analyzed using master equation techniques),
it is standard in modeling, as it typically brings 
out the salient features of the system with as few obscuring details from
the probabilistic model as possible. 
We then derive the functional forms 
of the following: $(i)$  Chord-specific inter-node distribution properties and
$(ii)$ for every outgoing pointer of a  Chord node, the probability 
that it is in any one of its possible states. 
This probability is different for each of the successor and finger pointers. 
 We then use this information to  predict other
quantities such as $(iii)$ the probability that the network gets disconnected, $(iv)$  lookup consistency (number of failed lookups), 
and $(v)$ lookup performance (latency).  
All quantities are computed as a function of the parameters 
involved and all results are verified by simulations.

%

\section{Related Work}
\label{sec:related}

Closest in spirit to our work is the informal derivation in the original Chord paper \cite{chord:ton} of the
average number of timeouts encountered by a lookup. This quantity was approximated there
by the product of the average number of fingers used in a lookup 
times the probability that a given finger points to a departed node. Our
methodology not only allows us to derive the latter quantity 
systematically but also demonstrates how
this probability depends on which finger (or successor) is involved. Further
we are able to derive a precise relation relating this probability to
lookup performance and consistency accurately at any value of the system 
parameters. 

In the works of 
Aberer  \textit{et al.} \cite{aberer04tok} and Wang {\emph et al.}~\cite{wang03resilience}, DHTs
are analyzed under churn and the results are compared with simulations.
These analyses can also be classified as fluid models. However 
the main parameter is the probability that a random selected entry of a routing 
table is stale. In our analysis, we determine this quantity from system details 
and churn rates. 

A brief announcement of the results presented in this paper, has appeared earlier in \cite{KEAH1}.

\section{Our Implementation of Chord}
\label{sec:assum}
{\bf The Chord Ring.} The general philosophy of DHTs is to map a set of data items
onto a set of nodes where the insertion and lookup of items
is done using the unique keys that the items are given.
Chord's realization of that philosophy is as follows. Peers 
and data items are given unique keys (usually obtained by a cryptographic
hash of unique attribute like the IP address or public key for nodes, and filename 
or checksum for items) drawn from
a circular key space of size ${\cal K}$. The Chord system dictates that
the right place for storing an item is at the first alive node
whose key succeeds the key of the item. Since we refer to nodes
and items by their keys, the insertion and lookup
of items becomes a matter of locating the right ``successor''
of a key. All nodes have successor and predecessor pointers.
For $N$ nodes, using only the successor pointers to lookup items requires $\frac{1}{2}N$
hops on average. 

{\bf Fingers.} To reduce the average lookup path length, nodes keep ${\cal M} = \log_2{\cal K}$
pointers known as the ``fingers''. Using these fingers, 
a node can retrieve any key in $O(\log N)$ hops. 
The fingers of a node $n$ (where $n \in 0 \cdots
{\cal K}-1$) point to exponentially increasing distances of keys away from $n$.
That is, $\forall i \in 1..{\cal M}$, $n$ points to a node whose key is equal 
\minorchange{to} $n + 2^{i-1}$. 
We denote that key by $n.fin_i.start$. However, for a certain $i$, there might not be a node 
in the network whose key is equal to $n + 2^{i-1}$.  Therefore,
$n$ points to the first successor of $n + 2^{i-1}$ which we denote by 
$n.fin_i.node$. 

{\bf The Successor List} Moreover, each node keeps a list of the ${\cal S}=O(\log(N))$ immediate successors
as backups for its first successor. We use the notation $n.s$ to refer
to this list and $n.s_i$ to refer to the $i^{th}$ element in the list. 
Finally we use the notation $n.p$ to refer to the predecessor.

{\bf Stabilization, Churn \& Steady State.}
To keep the pointers up-to-date in the presence of churn, each node performs periodic stabilization
of its successors and fingers. In our analysis, we define
$\lambda_j$ as the rate of joins per node, $\lambda_f$ the rate of failures per node and $\lambda_s$ the
rate of stabilizations per node. 
The fraction of stabilizations which act on the successors
is $\alpha$, such that
the rate of successor stabilizations is
$\alpha\lambda_s$, 
and the rate of finger stabilizations is $(1-\alpha)\lambda_s$. In all that
follows, we impose the steady state condition
$\lambda_j=\lambda_f$ unless otherwise stated. Further it is useful to define $r \equiv
\frac{\lambda_s}{\lambda_f}$ which is the relevant ratio on which all
the quantities we are interested in will depend, e.g, $r=50$ means
that a join/fail event takes place every half an hour for a
stabilization which takes place once every $36$ seconds.
Throughout the paper we will use the terms $\lambda_j \minorchange{N} 
\Delta t$, $\lambda_f \minorchange{N} \Delta t$, $\alpha \lambda_s \minorchange{N} \Delta t$ and $(1-\alpha) \lambda_s \minorchange{N} \Delta t$ to denote 
the respective probabilities that a join, failure, 
a successor stabilization, or
a finger stabilization take place \minorchange{anywhere on the ring} during a 
micro period of time of length $\Delta t$.

{\bf Parameters.} The parameters of the problem are hence: ${\cal K}$, $N$, $\alpha$ and $r$. 
All relevant measurable quantities should be entirely expressible in terms of these parameters. 

{\bf Simulation}
Since we are collecting statistics like the probability of a particular finger pointer to be wrong, we need to repeat each experiment $100$ times before obtaining well-averaged results. 
The total simulation sequential real time for obtaining the results of this paper was about $1800$ hours that was parallelized on a cluster of $14$ nodes where we had $N=1000$, ${\cal K}=2^{20}$, ${\cal S}=6$, $200 \leq r \leq 2000$
and $0.25 \leq \alpha \leq 0.75$.

While the main outlines of the chord protocol are provided by its authors
in \cite{chord:ton}, an exact analysis necessitates the provision of a deeper level of
detail and adopted assumptions which we provide in the following subsections.

\subsection{Joins, Failures \& Ring Stabilization}
\label{sec:stab}
{\bf Initialization.}
Initially, a node knows its key and at least one node with key $c$ that already exists in the network and is alive.
The knowledge of such a node is assumed to be acquired through some out-of-band method. 
The predecessor $p$, successors ($s_{1..{\cal S}}$) and fingers ($fin_{1..{\cal M}}.node$) 
are all assigned to $nil$. 

{\bf Joins} (Fig. \ref{fig:fixsucc}). A new node $n$ joins by looking up its successor using the initial 
random contact node $c$. It also starts its first stabilization of the successors and initializes its fingers.

{\bf Stabilization of Successors} (Fig. \ref{fig:fixsucc}). 
The function {\it fixSuccessors} is triggered periodically with rate $\alpha\lambda_s$.
A node $n$ tells its first alive successor $y$ that it believes itself 
to be $y$'s predecessor and expects as an answer $y$'s predecessor $y.p$
and successors $y.s$. The response of $y$ can lead to three actions:\\
{\it Case A}. 
Some node exists between $n$ and $y$ 
(\textit{i.e.}, $n$'s belief is wrong), so $n$ prepends $y.p$
to its successor list as a first successor and retries \emph{fixSuccessors}.\\
{\it Case B}. 
$y$ confirms $n$'s belief and informs $n$ of $y$'s old predecessor $y.p$. 
Therefore $n$ considers $y.p$ as an alternative/initial predecessor for $n$. 
Finally, $n$ reconciles its successor list with $y.s$.\\
{\it Case C}. 
$y$ agrees that $n$ is its predecessor and the only task of $n$ is to update
its successor list by reconciling it with $y.s$.

By calling \emph{iThinkIamYourPred} (Fig. \ref{fig:fixsucc}), some node $x$ informs $n$ that it believes itself to be $n$'s predecessor. If $n$'s predecessor $p$ is not alive or $nil$, then $n$ accepts $x$ as a predecessor and informs $x$ about this agreement by returning $x$. Alternatively, if $n$'s predecessor $p$  is alive (discovering that will be explained shortly in section \ref{sec:fail}), then there are two possibilities: The first is that $x$ is in the region between $n$ and its current predecessor $p$, therefore $n$ should accept $x$ as a new predecessor and inform $x$ about its old predecessor. The second  is that $p$ is already pointing to $x$ so the state is correct at both parties and $n$ confirms that to $x$ by informing it that $x$ is the predecessor of $n$. In all cases the function returns a predecessor and a successor list.

The function \emph{firstAliveSuccessor} (Fig. \ref{fig:fixsucc}) iterates through the successor list. In each iteration, if the first successor $s_1$ is alive, it is returned.  Otherwise, the dead successor is dropped from the list and nil is appended to
the end of the list. If the first successor is \emph{nil} this means that all immediate 
successors are dead and that the ring is disconnected.

\begin{figure}[t]
\fbox{
\begin{minipage}{0.95\columnwidth}
\subbegin{$n$.\textbf{join}($c$)}{join}
\SP $s_1$ = $c$.findSuccessor($n$)	
\SP fixSuccessors()
\SP initFingers($s_1$)
\subend

\footnotesize
\subbegin{$n$.\textbf{fixSuccessors}()}{fixsucc}
\SP			$y=$ firstAliveSuccessor()
\SP			$\{y.p,y.s\}$ = $y$.iThinkIamYourPred($n$)
\SP     if ($y.p \in (me,y)$) \Comment{Case A}
\ST			prepend($y.p$)	
\ST			fixSuccessors() 
\SP     elsif ($y.p \in (y,me)$) \Comment{Case B}
\ST			considerANewPred($y.p$)
\ST			reconcilce($y.s$)
\SP			else 	\Comment{Case C: $y.p == me$}
\ST			reconcile($y.s$)
\subend


\begin{tabular}{l|l}
\begin{minipage}{0.4\columnwidth}
\subbegin{$n$.\textbf{firstAliveSuccessor}()}{fas}
\SP			while (true) 
\ST				if ($s_1 == nil$)
\STT				\Comment{Broken Ring!!}
\ST				if (isAlive($s_1$))
\STT				return ($s_1$)
\ST				$\forall i \in 1..({\cal S}-1)$
\STT					$s_i = s_{i+1}$
\ST				$s_{S} = nil$
\subend			
\end{minipage}
&
\begin{minipage}{0.6\columnwidth}
\subbegin{$n$.\textbf{iThinkIAmYourPred}($x$)}{thinkpred}
\SP			if ((isNotAlive($p$) or ($p == nil$))
\ST					$p=x$
\ST					return($\{s,x\}$)
\SP			if ($x \in (p,me)$)
\ST				$oldp = p$ 
\ST				$p = x$
\ST				return($\{s,oldp\}$)
\SP			else
\ST				return($\{s,p\}$)
\subend
\end{minipage}
\\
\end{tabular}

\begin{tabular}{l|l|l}
\begin{minipage}{0.34\columnwidth}
\subbegin{$n$.\textbf{considerANewPred}($x$)}{consider}
\SP			if (isNotAlive($p$)  
\ST				 or ($p == nil$)  
\ST        or ($x \in (p,n)$))
\STT			$p = x$
\subend
\end{minipage}
&
\begin{minipage}{0.3\columnwidth}
\subbegin{$n$.\textbf{reconcile}($s'$)}{reconcile}
\SP		for $i = 1..({\cal S}-1)$	
\ST			$s_{i+1} = s'_{i}$
\subend
\end{minipage}
&
\begin{minipage}{0.31\columnwidth}
\subbegin{$n$.\textbf{prepend}($y$)}{prepend}
\SP		for $i={\cal S}..2$	
\ST			$s_{i} = s_{i-1}$
\SP		$s_1=y$
\subend
\end{minipage}
\\
\end{tabular}

\end{minipage}
}
\caption{Joins and Ring Stabilization Algorithms.}
\label{fig:fixsucc}
\end{figure}

\subsection{Lookups and Stabilization of Fingers}
\label{sec:fingers}

{\bf Stabilization of Fingers} (Fig. \ref{fig:fixfingers}).
Stabilization of fingers occurs at a rate $(1-\alpha)\lambda_s$.
Each time the \emph{fixFingers} function is triggered, a random
finger $fin_i$ is chosen and a lookup for $fin_i.start$ is
performed and the result is used to update $fin_i.node$.

\begin{figure}[t]
\flushleft
\fbox{
\begin{minipage}{0.95\columnwidth}
\subbegin{$n$.\textbf{initFingers}($s_1$)}{initfingers}
\SP $f' = s_1.f$
\SP   $\forall i \in 1..{\cal M}$ s.th. ($fin_i.start \in (n,s_1]$),
\ST			 $fin_i.node = s_1$
\SP		$\forall j \in 1..{\cal M}$ s.th. ($fin_j.start \notin (n,s_1]$),  
\ST		   $fin_j.node = $localSuccessor$(f',fin_j.start)$
\subend
\begin{tabular}{c|c}
\begin{minipage}{0.35\columnwidth}
\subbegin{$n$.\textbf{localSuccessor}($f$,$k$)}{localsucc}
\SP	for $i=1..{\cal M}$
\ST		if ($k \in (n,fin_i]$) 
\STT			return($fin_i$)
\SP	return(nil)
\subend
\end{minipage}
& \begin{minipage}{0.55\columnwidth}
\subbegin{$n$.\textbf{fixFingers}($k$)}{fixfingers}
\SP 	$1 \leq i =$ random() $\leq {\cal M}$ 
\SP		$fin_i$.node = findSuccessor($fin_i$.start)
\subend
\end{minipage}
\\
\end{tabular}

\end{minipage}
}
\caption{Initialization and Stabilization of Fingers.}
\label{fig:fixfingers}
\end{figure}

\begin{figure}[ht]
\fbox{
\begin{minipage}{0.95\columnwidth}
\subbegin{$n$.\textbf{findSuccessor}($k$)}{findsucc}
\SP		\Comment{Case A: $k$ is exactly equal to $n$}
\SP		if ($k$ == $n$)
\ST				return($n$)
\SP		\Comment{Case B: $k$ is between $n$ and $s_1$}
\SP		if ($k \in (n, s_1]$)
\ST			return(firstAliveSuccessorNoChange());
\SP		\Comment{Case C: Forward to the lookup to}
\SP		\Comment{the closest preceding alive finger}
\SP		$cpf$ = closestAlivePrecedingFinger($k$);
\SP		if ($cpf == nil$)
\ST			$y$ = firstAliveSuccessorNoChange();
\ST			if ($k \in (n, y]$)
\STT				return($y$);		
\ST   	$cpf$ = closestAlivePrecedingSucc(k);
\ST			return($cpf$.findSuccessor(k))
\SP		else
\ST			return ($cpf$.findSuccessor(k));
\subend
\subbegin{$n$.\textbf{firstAliveSuccessorNoChange}()}{fasnc}
\SP			$i = 1$
\SP			while (true) 
\ST				if ($s_i == nil$)
\STT				//Broken Ring!!
\ST				if (isAlive($s_i$))
\STT				return ($s_i$)
\ST				$i++$
\subend
\begin{tabular}{c|c}
\begin{minipage}{0.475\columnwidth}
\subbegin{$n$.\textbf{closestAlivePrecedingFinger}($k$)}{cpf}
\SP	for $i={\cal M}..1$
\ST		if (($fin_i \in (n, k)$)
\STT		  and ($fin_i \neq nil$)
\STT			and isAlive($fin_i$))
\STTT					return($fin_i$)
\SP	return(nil)
\subend
\end{minipage}
&
\begin{minipage}{0.425\columnwidth}
\subbegin{$n$.\textbf{closestAlivePrecedingSucc}($k$)}{cps}
\SP		for $i={\cal S}..1$
\ST		if (($s_i \in (n, k)$) 
\STT		and ($s_i \neq nil$)
\STT		and isAlive($s_i$))
\STTT					return($s_i$)
\SP		return(cpf)
\subend
\end{minipage}
\\
\end{tabular}

\end{minipage}
}
\caption{The Lookup Algorithm.}
\label{fig:findsucc}
\end{figure}


{\bf Initialization of Fingers} (Fig. \ref{fig:fixfingers}).
After having initialized its first successor $s_1$, 
a node $n$ sets all fingers with starts between $n$ and $s_1$ to $s_1$.
The rest of the fingers are initialized by taking a copy of the finger
table of $s_1$ and finding an approximate successor to every finger from that finger
table.

{\bf Lookups} (Fig. \ref{fig:findsucc}). A lookup operation is a fundamental operation that is used to
find the successor of a key. It is used by many other routines and
its performance and consistency are the main quantities of interest
in the evaluation of any DHT. 
A node $n$ looking up the successor of $k$ runs the \emph{findSuccessor}
algorithm which can lead to the following cases:

{\it Case A.} If $k$ is equal to $n$ then $n$ is trivially the
successor of $k$.

{\it Case B.} If $k \in (n,s_1]$ then  $n$ has found the
successor of $k$, but it could be that $s_1$ has failed
and $n$ has not yet discovered this. However, entries in 
the successor list can act as backups for the first successor. 
Therefore, the first alive successor of $n$ is
the successor of $k$. Note that, in this case,  while we try to
find the first alive successor, we do not change the entries
in the successor list. This is mainly because, to simplify the analysis,
we want the successor list to be changed at a fixed rate rate $\alpha\lambda_s$
only by the \emph{fixSuccessors} function.

{\it Case C.} The lookup should be forwarded to a node closer to $k$, namely the closest 
alive finger preceding $k$ in $n$'s finger table. The call to the function
\emph{closestAlivePrecedingFinger} returns such a node if possible and the lookup
is forwarded to it. However, it could be the case that all alive preceding fingers 
to $k$ are dead. In that case, we need to use the successor list as a last
resort for the lookup. Therefore, we locate the first alive successor $y$ and
if $k \in (n,y]$ then $y$ is the successor of $k$. Otherwise, we locate
the closest alive preceding successor to $k$ and forward the lookup to it. 

\subsection{Failures}
\label{sec:fail}
Throughout the code we use the call $isAlive$ and $isNotAlive$. A simple interpretation
of those routines would be to equate them to a performance of a ping. However, a correct
implementation for them is that they are discovered by performing the operation required.
For instance, a call to $firstAliveSuccesor$ in Fig. \ref{fig:fixsucc} is performed to 
retrieve a node $y$ and then call $y.iThinkIamYourPred$, so alternatively the first alive
successor could be discovered by iterating on the successor list and calling $iThinkIamYourPred$.

\section{The Analysis}
\label{sec:internode}
\subsection{Distributional Properties of Inter-Node Distances}
In this section we will assume that all keys are populated by peers
with independent and equal probability, and, furthermore, that
this probability does not change with time.
The first condition is a natural consequence
of peers joining and leaving/failing
independently. The last condition, on the other hand, 
does not hold strictly since the number
of peers present under churn is a fluctuating quantity,
Nevertheless, it can be expected to hold to good accuracy 
in sufficiently large systems.
A detailed analysis along these lines will be given elsewhere.

\begin{definition}
\label{def:int}
Given two keys $u, v \in \{0...{\cal K}-1\}$, the ``distance'' 
between them is $u - v $ (with modulo-{$\cal K$} arithmetic). 
We interchangeably say that $u$ and $v$ form an 
``interval'' of length  $u - v $. Hence
the number of keys 
{\it inside} an interval of length $\ell$ is $\ell-1$ keys. 
\end{definition}


\begin{property}
\label{prop:px} 


The probability $P(x)$ 
of finding an interval of length $x$ is:
$P(x) = \rho^{x-1}(1-\rho)$ where $\rho = \frac{{\cal K}-N}{\cal K}$.
\end{property}
Under the stated conditions, each \minorchange{key} will be populated
with the same probability $\frac{N}{\cal K}=1-\rho$, \minorchange{for $N<<K$}.
An interval of length $x$ then involves $x-1$ consecutive
unpopulated  \minorchange{keys}, and then one populated \minorchange{key}, which explains the formula.

\begin{figure}[t]
	\centering
	\includegraphics{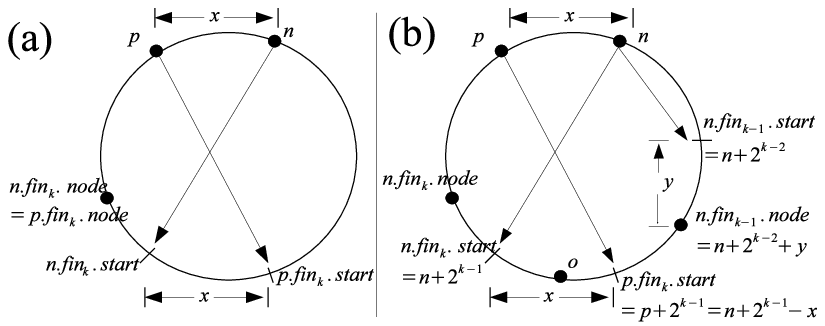}
	\caption{(a) Case when $n$ and $p$ have the same value of $fin_k.node$.
	         (b) Case where a newly joined node $p$ copies the $k^{th}$ entry
of its successor node $n$ as the best approximation for its own $k^{th}$ entry 
(by the join protocol). In this case, there could be a node $o$ which
is the 'correct' entry for $p.fin_k.node$. However, since $p$ is newly joined, the only information it has access to is the finger table of $n$.}
	\label{fig:props}
\end{figure}

We now derive some properties of this distribution 
which will be used in the ensuing analysis. 
\begin{property}
\label{prop:ab}
For any two keys $u$ and $v$, where $v=u+x$, let $b_i$ be the probability
that the first node encountered in between these two keys is at $u+i$ 
(where $0 \leq i < x$).
Then $b_i \equiv {\rho^{i}(1-\rho)}$. 
The probability that there is definitely at least one node between $u$ and $v$ is: $a(x)\equiv {1-\rho^x}$. 
Hence the conditional probability that the first node is at a distance $i$ {\it given} that
there is at least one node in the interval is $ bc(i,x)\equiv b(i)/a(x)$.
\end{property}

\begin{property}
\label{prop:share}
The probability that a node and at least one of its immediate predecessors 
share the same $k^{th}$ finger
is $p_1(k)\equiv \frac{\rho}{1+\rho} (1-\rho^{2^k-2})$. 
The explanation for this property goes as follows. 
If the distance between node $n$ and its predecessor $p$ is $x$, the distance between 
$n.fin_k$.\emph{start} and $p.fin_k$.\emph{start} is also $x$ (see Fig. \ref{fig:props}(a)). If there is no node
in between $n.fin_k$.\emph{start} and $p.fin_k$.\emph{start} then $n.fin_k$.\emph{node} and $p.fin_k$.\emph{node}
will share the  same value. From Property~\ref{prop:px}, 
the probability that the distance between $n$ and $p$ is $x$ 
is $\rho^{x-1}(1-\rho)$. However, $x$ has to be less than $2^{k-1}$, otherwise 
$p.fin_k$.\emph{node} will be equal to $n$.  The probability that no node 
exists between $n.fin_k$.\emph{start} and $p.fin_k$.\emph{start}
is $\rho^x$ (by Property \ref{prop:ab}). 
Therefore the probability that the $n.fin_k$.\emph{node} and 
$p.fin_k$.\emph{node} share the same value is:
$\sum_{x=1}^{2^{k-1}-1} \rho^{x-1}(1-\rho)\rho^x  =  \frac{\rho}{1+\rho} (1-\rho^{2^k-2})$.
It is straightforward (though tedious) to
derive similar expressions for $p_2(k)$ the probability that a node and 
at least {\it two} of its immediate predecessors share the same $k^{th}$ finger,
$p_3(k)$ and so on.
\end{property}


\begin{property}
\label{prop:copy}
We can similarly assess the probability that the join 
protocol (see Section~\ref{sec:fingers})
results in further replication of the $k^{th}$ pointer. Let us define the probability
$p_{\minorchange{\it join}} (i,k)$ as the probability that a newly joined node, chooses 
the $i^{th}$ entry of its successor's finger table for its own $k^{th}$ entry. Note 
that this is unambiguous even in the case that the successor's $i^{th}$
entry is repeated. All we
are asking is, when is the $k^{th}$ entry of the new joinee the same as the $i^{th}$
entry of the successor? Clearly $ i \leq k$. 
In fact for the larger fingers, we only need to consider 
$p_{\minorchange{\it join}} (k,k)$, since $p_{\minorchange{\it join}}(i,k) \sim 0 $ for $i<k$.
 Using the interval distribution we find, for large $k$,
$p_{\mathrm join}(k,k) \sim \rho (1-\rho^{2^{k-2} -2}) + (1-\rho) (1-\rho^{2^{k-2}-2}) -(1-\rho) \rho (2^{k-2} -2) \rho^{2^{k-2}-3} $. This function goes to $1$ for large $k$.
\end{property}

We can also analogously compute $p_{\minorchange{\it join}}(i,k)$ for any $i$.
The only trick here is to estimate the probability that
starting from $i$, the last {\it distinct} entry of $n$'s finger table
{\it does not} give $p$ a better choice for its $k_{th}$ entry.
This can again readily be computed using property \ref{prop:ab}, but we
do not do the computation here since for our purposes $p_{\minorchange{\it join}}(k,k)$
suffices.


\subsection{Successor Pointers}
\begin{figure}
	\centering
	\includegraphics[width=9cm, height=7cm]{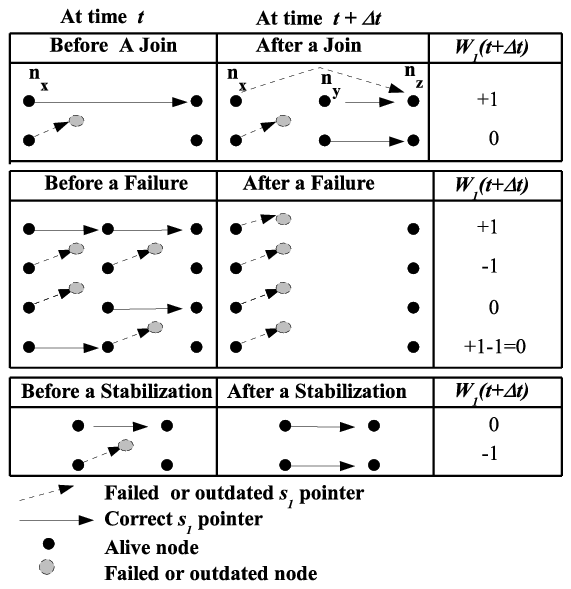}
	\caption{Changes in $W_1$, the number of wrong  (failed or outdated) $s_1$ pointers, due to joins, failures and stabilizations.}
	\label{fig:w1-trans}
\end{figure}
%
We now turn to estimating various quantities of interest for Chord.
In all that follows we will evaluate various {\it average} quantities, 
as a function of the parameters. To do this we need
to understand how the dynamical evolution of the system affects these
quantities.

In the case of Chord, we only need to consider  one of 
three kinds of events happening at any micro-instant: a join, a failure
or a stabilization. One assumption made in the following
is that such a micro-instant of time exists, or in other words, that
we can divide time till we have an interval small enough that 
in this interval, only one of these three processes occurs 
\minorchange{anywhere in the system}. 
Implicit in this is the assumption that a stabilization
(either of successors or fingers) is done faster than the
time-scales over which joins and fails occur. 

Another aspect of this system which simplifies analysis is that
successor pointers of adjacent nodes are independent of each other.
That is, the state of the first successor pointer of a given node
does not affect the state of the first successor pointer of either its 
predecessor or its successor. The same logic also works for the 
state of the second successor pointers of adjacent nodes and so on.
On the other hand, the state of the second successor pointer
of a node is clearly related to the state of its first successor 
pointer as well the state of the first successor pointer of the successor. 
This is taken into account in the analysis of second and higher successor 
pointers. In characterizing the states of higher successors, 
we look for the leading order behavior in terms of the 
parameter $r$.

Consider first the successor pointers.
Let $w_k(r,\alpha)$ denote the fraction of nodes having 
a \emph{wrong} $k^{th}$ successor pointer and 
$d_{k}(r,\alpha)$ the fraction of nodes having 
a \emph{failed} successor pointer.
Also, let $W_k(r,\alpha)$ be 
the number of nodes having 
a \emph{wrong} $k^{th}$ successor pointer and 
$D_{k}(r,\alpha)$ the number of nodes having 
a \emph{failed} successor pointer.
A \emph{failed} pointer is one
which points to a departed node while
a \emph{wrong} pointer points either to an
incorrect node (alive but not correct) or a dead one. 
As we will see, both these quantities play a role 
in predicting lookup consistency and lookup length.

By the protocol for stabilizing successors in Chord, a node periodically contacts its first successor, possibly correcting it and reconciling with its successor list. Therefore, the number of wrong $k^{th}$ successor pointers are not independent quantities but depend on the number of wrong first successor pointers. 


%


\begin{table}[t]
\caption{Gain and loss terms for $W_1(r,\alpha)$: the number of wrong first successors
as a function of $r$ and $\alpha$.} 
\label{tab:wrong}
	\centering
		\begin{tabular}{|l|l|} \hline
		Change in $W_1(r,\alpha)$	&  \minorchange{Probability of Occurrence}   \\ 
		$W_1(t+\Delta t) = W_1(t)+1$ & $c_{1.1}=(\lambda_j \minorchange{N} \Delta t) (1-w_1)$ \\ 
		$W_1(t+\Delta t) = W_1(t)+1$ & $c_{1.2}=\lambda_f \minorchange{N} (1-w_1)^2   \Delta t$ \\ 
		$W_1(t+\Delta t) = W_1(t)-1$ & $c_{1.3}=\lambda_f \minorchange{N} w_1^2   \Delta t $ \\ 
		$W_1(t+\Delta t) = W_1(t)-1$ & $c_{1.4}=\alpha\lambda_s \minorchange{N} w_1   \Delta t $\\ 
		$W_1(t+\Delta t) = W_1(t)$ & $1 - (c_{1.1} + c_{1.2} + c_{1.3} + c_{1.4})$\\ 
\hline
		\end{tabular}
\end{table}

We write an equation for $W_1(r,\alpha)$ by accounting  for all the events that can change it in a micro event of time $\Delta t$. An illustration of the different cases in which changes in $W_1$ take place due to joins, failures and stabilizations is provided in Fig. \ref{fig:w1-trans}. In some cases $W_1$ increases/decreases while in others it stays unchanged. For each
increase/decrease, Table \ref{tab:wrong} provides the corresponding 
\minorchange{probabilities}. 

By our implementation of the join protocol, a new node $n_y$, joining between two nodes $n_x$ and $n_z$, always has a correct $s_1$ pointer after the join. However the state of $n_x.s_1$ before the join makes a difference. If $n_x.s_1$ was correct (pointing to $n_z$) before the join, then after the join it will be wrong and therefore $W_1$ increases by $1$. If $n_x.s_1$ was wrong before the join, then it will remain wrong after the join and $W_1$ is unaffected. Thus, we need to account for the former case only. The probability that $n_x.s_1$ is correct is $1-w_1$ and term $c_{1.1}$ follows from this. 

For failures, we have $4$ cases. To illustrate them we use nodes $n_x$, $n_y$, $n_z$ and assume that $n_y$ is going to fail.
First, if both $n_x.s_1$ and $n_y.s_1$ were correct, then the failure of $n_y$ will make $n_x.s_1$ wrong and hence $W_1$ increases by $1$. Second, if $n_x.s_1$ and $n_y.s_1$ were both wrong, then the failure of $n_y$ will decrease $W_1$ by one,
since one wrong pointer disappears. Third, if $n_x.s_1$ was wrong
and $n_y.s_1$ was correct, then $W_1$ is unaffected. Fourth, if $n_x.s_1$ was correct and $n_y.s_1$ was wrong, then the wrong pointer of $n_y$ disappears and $n_x.s_1$ becomes wrong, therefore $W_1$ is unaffected. For the first case to happen, we need to pick two nodes with correct pointers, the probability of this is $(1-w_1)^2$. For the second case to happen, we need to pick two nodes with wrong pointers, the probability of this is $w^2_1$. From these probabilities follow the terms $c_{1.2}$ and $c_{1.3}$.

Finally, a successor stabilization does not affect $W_1$, unless the stabilizing node had a wrong pointer. The probability of picking such a node is $w_1$. From this follows the term $c_{1.4}$. 

Hence the equation for $W_1(r,\alpha)$ is: 
\begin{equation}
\frac{d W_1}{\minorchange{N} dt}= \lambda_j (1-w_1) + \lambda_f (1-w_1)^2  - \lambda_f w_1^2 - \alpha\lambda_s w_1    \nonumber
\end{equation}
Solving for $w_1$ in the steady state and putting $\lambda_j=\lambda_f$, we get:
\begin{equation}
w_1(r,\alpha) = \frac{2}{3+r\alpha} \approx \frac{2}{r\alpha}
\end{equation}

\begin{figure}
	\centering
		\includegraphics[height=8cm, angle=270]{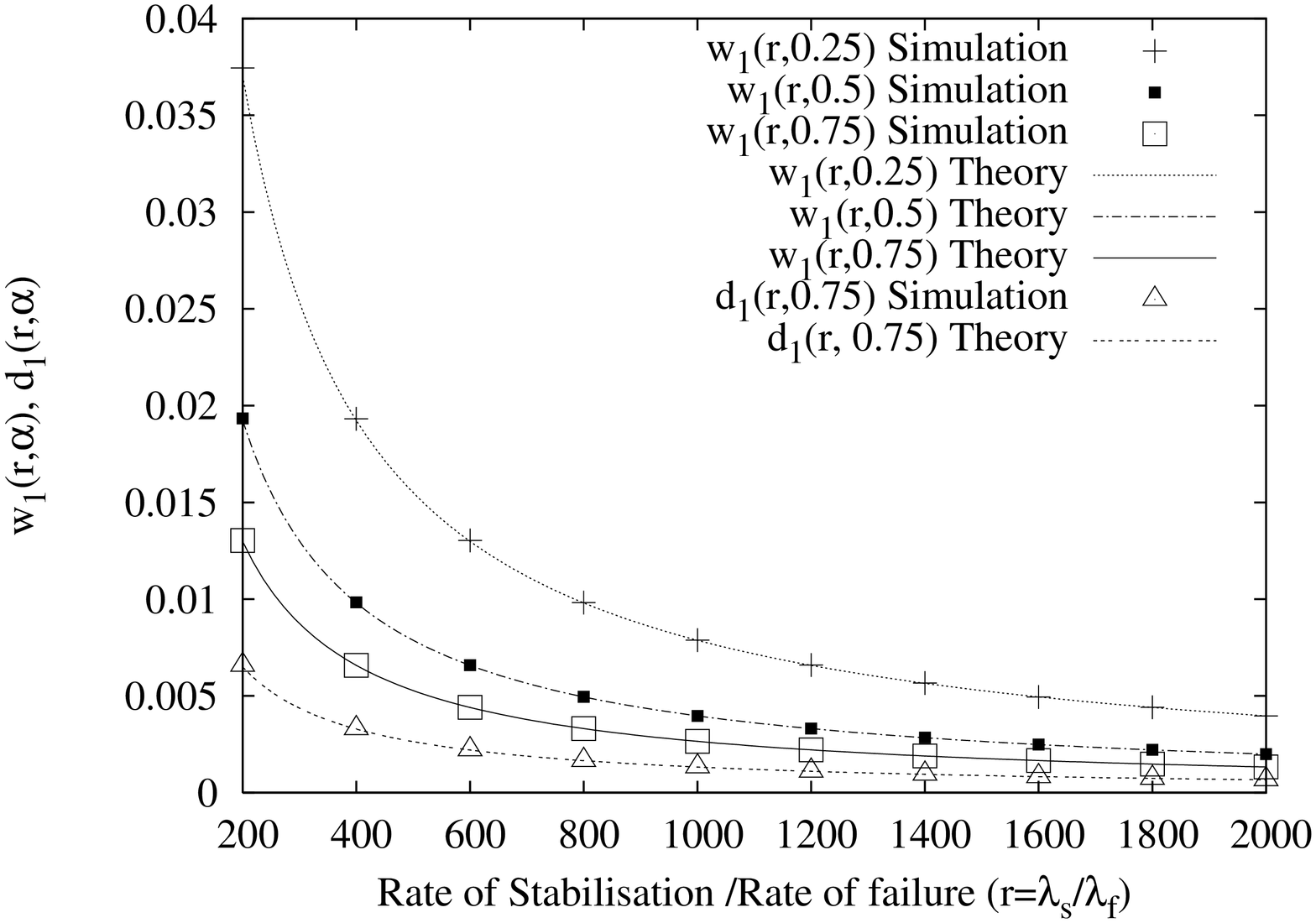}
	\caption{Theory and simulation for the probability of wrong $1^{st}$ successor $w_1(r,\alpha)$ and failed $1^{st}$ successor $d_1(r,\alpha)$.}
	\label{fig:wi}
\end{figure}

This expression matches well with the simulation results as shown in Fig. 
\ref{fig:wi}. 
$d_1(r,\alpha)$ is then $ \approx \frac{1}{2}w_1(r,\alpha)$
since when $\lambda_j=\lambda_f$, about half the number of wrong pointers
are incorrect and about half point to dead nodes. 
Thus $ d_1(r,\alpha) \approx \frac{1}{r\alpha}$ which
also matches well the simulations as shown in Fig. \ref{fig:wi}. 

The fraction of wrong second successors can be estimated in an analogous manner. 
Consider, for a node $n$, the possible states of 
the successor, $n.s_1$, the successor of the successor,
$*(n.s_1).s_1$, and
the second successor, $n.s_2$.
In a fully correct state, 
$*(n.s_1).s_1$ and $n.s_2$ of course point to the same node.
If in such a state either $n.s_1$ or $*(n.s_1).s_1$
becomes incorrect through the action of a join or a failure, then
 $n.s_2$ is also incorrect. On the other hand,  $n.s_2$
cannot be corrected by the stabilization protocol
unless both $n.s_1$ and $*(n.s_1).s_1$ are both already corrected.
Hence,  $n.s_2$ is wrong if either $n.s_1$  or $*(n.s_1).s_1$ are
wrong, and also if both $n.s_1$  and $*(n.s_1).s_1$ are correct,
but  $n.s_2$ has not yet been corrected.
If the number of such non-stabilized
configurations is $N_2$ and the fraction is $n_2$, we have
\begin{equation}
\label{eq:w2-equation}
w_2 = 2w_1 - w_1^2 + n_2
\end{equation}

To estimate $n_2$ we consider how these configurations might be 
gained or lost.  The gain term arises
from stabilizations of configurations
where $n.s_1$ is correct but $*(n.s_1).s_1$  is wrong.
A stabilization performed by node $n.s_1$ then
results in the gain of a $N_2$ configuration.
On the other hand, non-stabilized configurations are lost either
by a stabilization performed by node $n$ (when it gets the correct 
successor list from its successor and hence corrects $n.s_2$), 
or by corrupting either  $n.s_1$ or $*(n.s_1).s_1$ 
(by a join or failure).  The latter possibility
gives terms of order $\frac{1}{r^2}$ and we can ignore
it in the limit \minorchange{that} stabilizations happens on
a much faster time scale than joins and failures  (\textit{i.e.},
$r$ much larger than unity). The equation for $N_2$ is hence
\begin{equation}
\label{eq:n2-equation}
\frac{dN_2}{dt} \approx
\alpha\lambda_s w_1 (1-w_1) - \alpha\lambda_s n_2
\end{equation}
which implies $n_2\approx w_1$ to order $\frac{1}{r}$.
Thus, we have $w_2 \approx \frac{6}{r}$.

\begin{figure}
	\centering
		\includegraphics[height=8cm, angle=270]{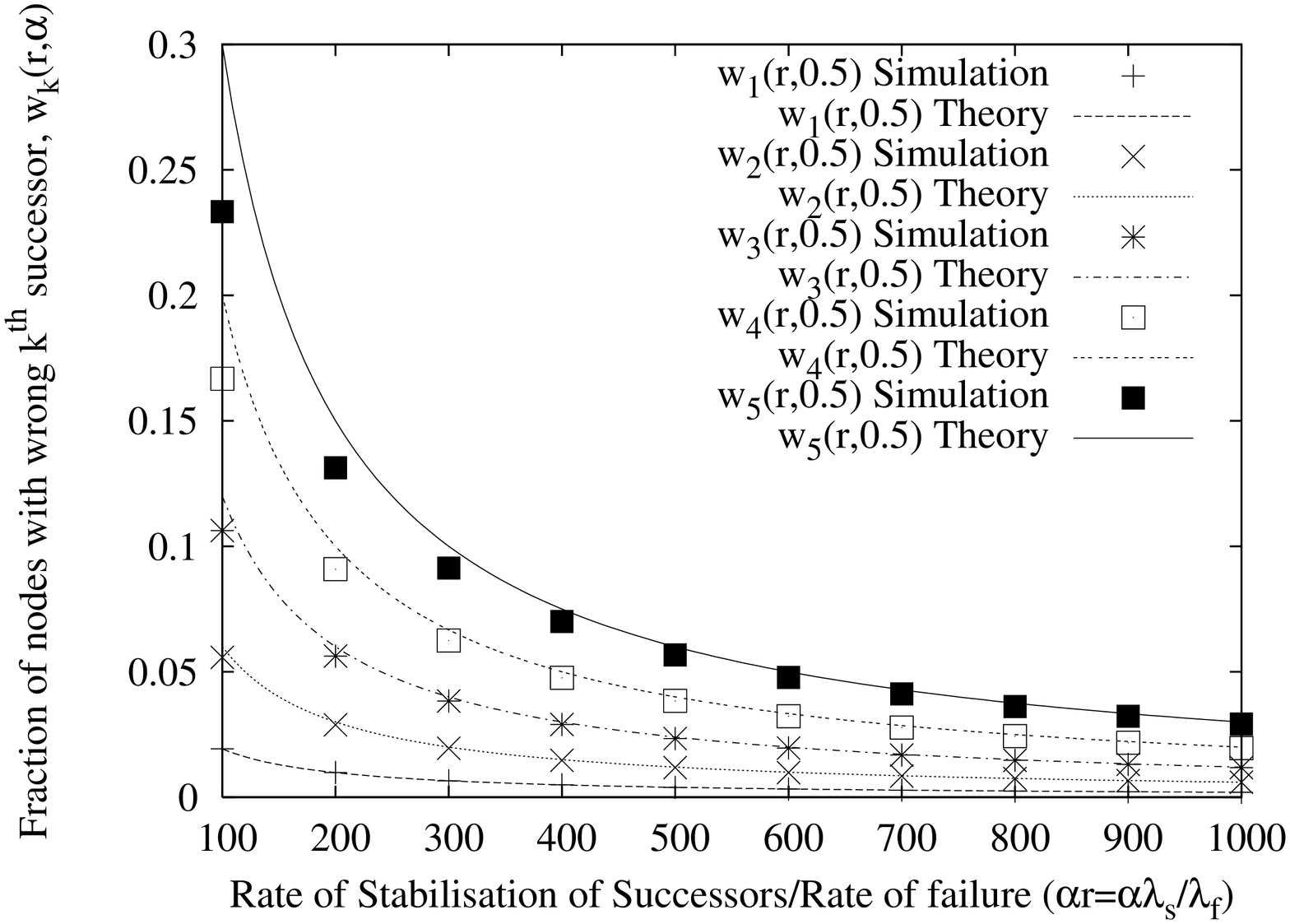}
	\caption{Theory and simulation for the probability of a wrong $k^{th}$ successor $w_k(r,\alpha)$.}
	\label{fig:wk}
\end{figure}

For higher successors we reason similarly by considering 
the state of the  ${k-1}^{st}$ successor pointer of node $n$, 
the successor pointer of the ${k-1}^{st}$ successor,
and the $k^{th}$ successor pointer of node $n$. 
We can write a recursion equation for $w_k$ the fraction of nodes with 
wrong $k^{th}$ successor pointer 
\begin{equation}
\label{eq:wk-equation}
w_k = w_1 + w_{k-1} - w_{k-1} w_1 + n_k
\end{equation}
where $n_k$ is the density of configurations where
the ${k-1}^{st}$ successor pointer of node $n$ and the first successor pointer
of the ${k-1}^{st}$ successor are both correct, but this information
has not yet been used to correct the $k^{th}$ successor pointer of node $n$.
If node $n$ does not as yet have the correct information about its
$k^{th}$ successor, that means that either all the nodes in between $n$ and its ${k-1}^{st}$  successor have the correct information but node $n$ has not as yet stabilized, or that the stabilization has propagated back from the ${k-1}^{st}$ successor
to  some node in between but not as yet to $n.s_1$.
To elaborate on this further, there is the case where the 
second successor pointer
of the ${k-2}^{nd}$  successor has not been corrected, then the case where
this has been done, but the third successor pointer of
the  ${k-3}^{rd}$ successor has not been corrected, and so on.
Each of these is analogous to $n_2$ and each occurs with density
$(1-w_{k-1})w_1$, if joins and failures are neglected compared
to stabilizations.
Hence, if to leading order in $\frac{1}{r}$ we have 
$w_k \sim \frac{c_k}{\alpha r}$, then
\begin{equation}
\label{eq:ck-equation}
c_k = c_{k-1} + k c_1
\end{equation}
which leads to 
\begin{equation}
\label{eq:wk-leading}
w_k \approx \frac{k(k+1)}{\alpha r}
\end{equation}.
We note that this expression obviously depends on the
details of the stabilization scheme, and is in principle 
only valid up to $k \sim \sqrt{r}$. 
As shown in Fig. \ref{fig:wk}, the agreement between
theory and simulation is still however quite reasonable
at $k=5$ and $r=100$. 

\subsection{Break-up (Network Disconnection) Probability}

\begin{table}[t]
\caption{Gain and loss terms for $N_{bu} (2,r, \alpha)$: 
the number of nodes with dead 
first {\em and} second successors.}
\label{tab:disconnect} 
	\centering
		\begin{tabular}{|l|l|} \hline
		Change in $N_{bu}(r,\alpha)$	&  \minorchange{Probability of Occurrence}   \\ 
		$N_{bu}(t+\Delta t) = N_{bu}(t)+1$ & $c_{2.1}=(\lambda_f \minorchange{N} \Delta t)d_1 (r, \alpha)$ \\ 
		$N_{bu}(t+\Delta t) = N_{bu}(t)+1$ & $c_{2.2}=\lambda_f \minorchange{N} \Delta t (1-d_1) d_2 $ \\ 
		$N_{bu}(t+\Delta t) = N_{bu}(t)-1$ & $c_{2.3}=\alpha \lambda_s \minorchange{N} \Delta t P_{bu}(2,r,\alpha) $ \\ 
		$N_{bu}(t+\Delta t) = N_{bu}(t)$ & $1 - (c_{2.1} + c_{2.2} + c_{2.3} )$\\ 
\hline
		\end{tabular}
\end{table}

We demonstrate below, how calculating $d_k(r, \alpha)$:
the fraction of nodes with dead $k^{th}$ pointers,
helps in estimating the probability that
the network gets disconnected for any value of $r$ and $\alpha$.
Let $P_{bu} (n, r,\alpha)$ be the probability that
$n$ consecutive nodes fail. If
$n={\cal S}$, the length of the successor list, then clearly the node
whose successor list this is, gets disconnected from the network 
and the network breaks up.
For the range of $r$ considered in Fig. \ref{fig:wi}, 
$P_{bu}({\cal S},r,\alpha) \sim 0$. However should we go lower, this 
starts becoming finite. The master equation analysis
introduced here can be used to estimate $P_{bu} (n,r,\alpha)$ 
for any $1\le n \le {\cal S}$. We 
indicate how this might be done by first considering the case $n=2$.
Let $N_{bu} (2,r,\alpha)$ be the number of configurations in which 
a node has both $s_1$ and $s_2$ dead and $P_{bu}(2,r,\alpha)$ be the 
fraction of such configurations. 
Table \ref{tab:disconnect} indicates how this is estimated
within the present framework.

\begin{figure}
	\centering
		\includegraphics[height=9cm, angle=270]{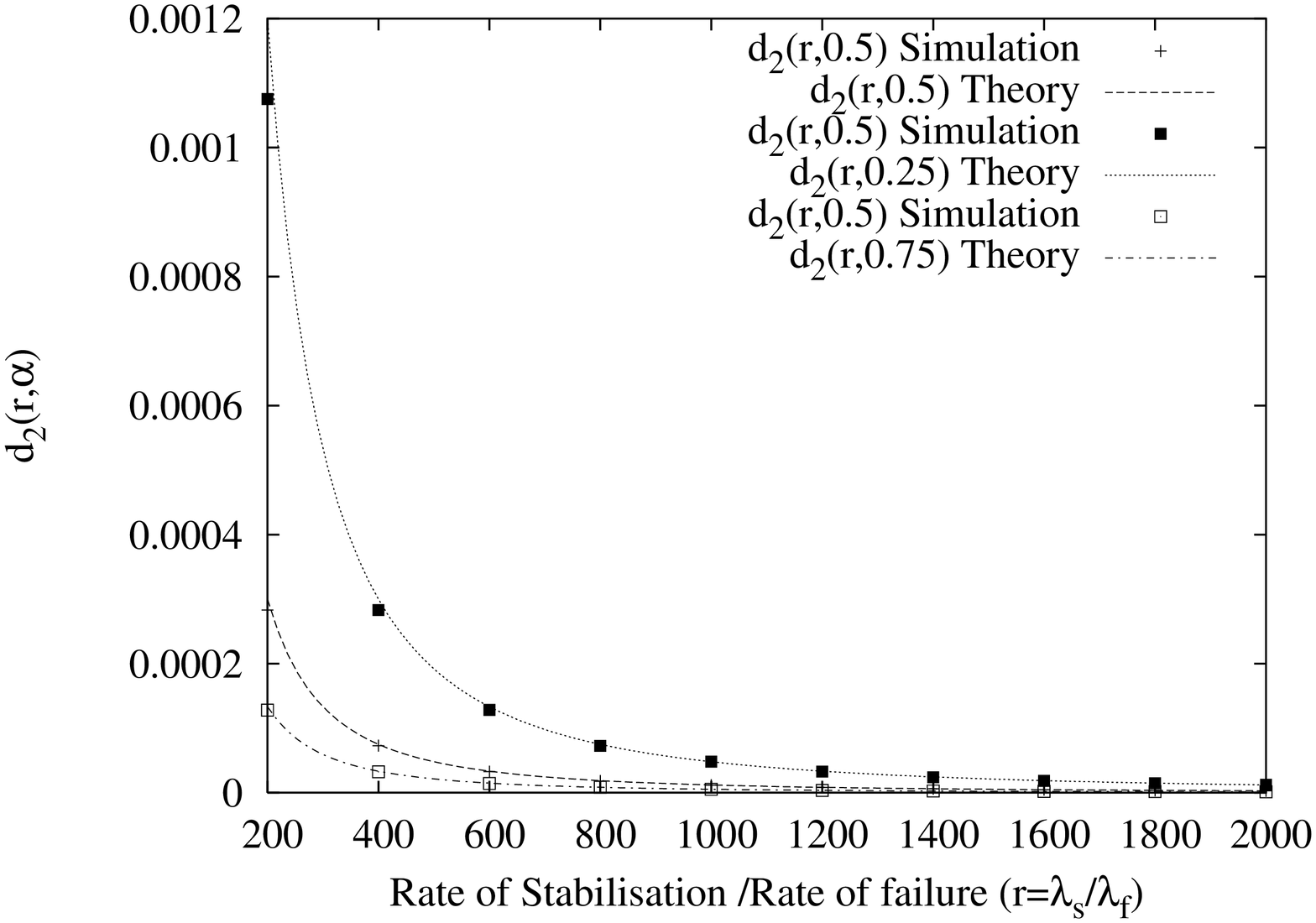}
	\caption{Theory and simulation for the probability of failure of the $2^{nd}$ successor, $d_2(r,\alpha)$.}
	\label{fig:d2}
\end{figure}

A join event does not affect this probability in any way. So we only need to
consider the effect of failures or stabilization events.
The term $c_{2.1}$ accounts for the situation when the 
{\em first} successor of a node is dead 
(which happens with probability $d_1 (r, \alpha)$ as explained above). 
A failure event can then kill its second successor as well and this happens
with probability $c_{2.1}$. The second term is the situation that the first
successor is alive (with probability $1-d_1$) but the second successor
is dead (with probability $d_2$). The logic used to estimate $d_2$
(or $d_k$ in general) is very similar to the reasoning 
we used to estimate the $w_k$'s. So we have 
\begin{equation}
\label{eq:dk-equation}
d_k = d_{1} + (k-1) d_1 = k d_1
\end{equation}
Thus the $k^{th}$ successor of a node is dead if the ${k-1}^{st}$
successor's successor is dead, or the ${k-1}^{st}$ successor's successor 
is not dead but the intermediate nodes think it is
because they haven't stabilized. 
Hence $d_2 \sim 2/\alpha r$. This estimate for $d_2$ matches the simulation results very well, as shown in Fig. \ref{fig:d2}.

Coming back to counting the gain and loss terms for $N_{bu}(2,r,\alpha)$, 
a stabilization event reduces 
the number of such configurations by one, if
the node doing the stabilization had such a configuration to begin with.

\begin{figure}
	\centering
		\includegraphics[height=9cm, angle=270]{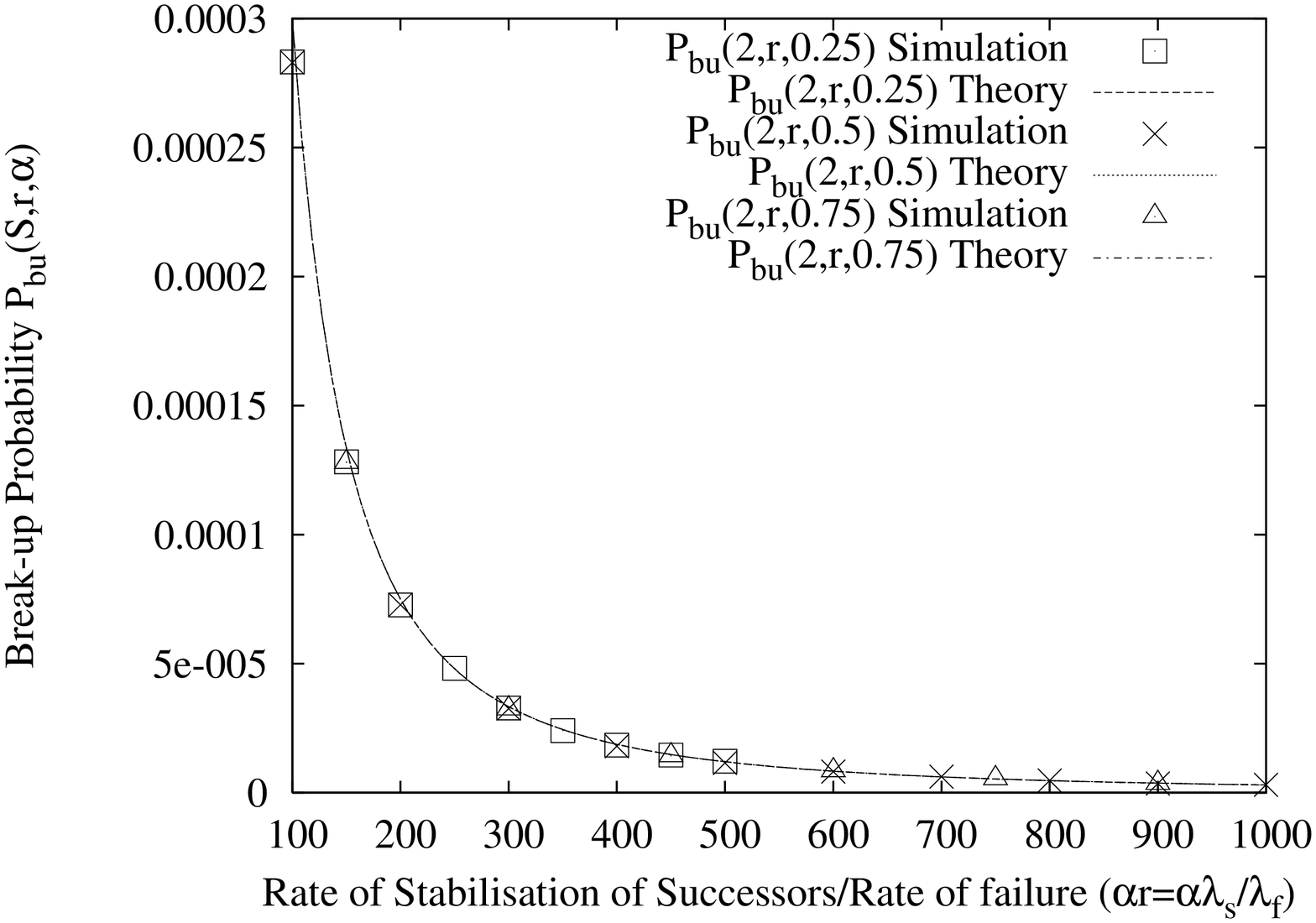}
	\caption{Theory and simulation for the break-up probability $P_{bu}(2, r, \alpha)$.}
	\label{fig:fos2}
\end{figure}

Solving the equation for $N_{bu} (2,r,\alpha)$, one hence obtains
that $P_{bu}(2,r,\alpha) \sim 3/(\alpha r)^2$. 
As Fig. \ref{fig:fos2} shows, this is a
precise estimate.

We can similarly estimate the probabilities for three consecutive nodes
failing, {\it etc}, and hence also the general disconnection
probability $P_{bu}({\cal S},r,\alpha)$. In fact
$P_{bu}({\cal S},r,\alpha)$ may be written in terms of the
$d_k(r, \alpha)$ as:
\begin{equation}
\label{eq:Pbu-equation}
P_{bu}({\cal S}) = ({{\cal S}-1})! \frac{\sum_{1}^{\cal S} d_i(r,\alpha)}{(\alpha r)^{{\cal S}-1}}
\end{equation}
The logic behind this equation is similar to that used for 
solving for $P_{bu}(2)$, namely that for ${\cal S}$ consecutive nodes to fail, any ${{\cal S}-1}$ of the  ${\cal S}$ nodes should have 
failed first, and then a failure event kills the remaining node.
(\ref{eq:Pbu-equation})  
is readily solved by substituting the values of the $d_k$'s to get
\begin{equation}
\label{eq:Pbu-solution}
P_{bu}({\cal S})= \frac{({{\cal S}+1})!}{2 (\alpha r)^{{\cal S}}}
\end{equation}

As mentioned above this is again correct only to leading order. Namely
there will be correction terms of the order $r^{{\minorchange{\cal S}} +1}$ which we haven't 
computed at this level of approximation. 
The Master Equation formalism 
thus affords the possibility of making a precise
prediction for when the system runs the danger of 
getting disconnected, as a function of the parameters.

\begin{figure}
        \centering
                \includegraphics[height=8cm, angle=270]{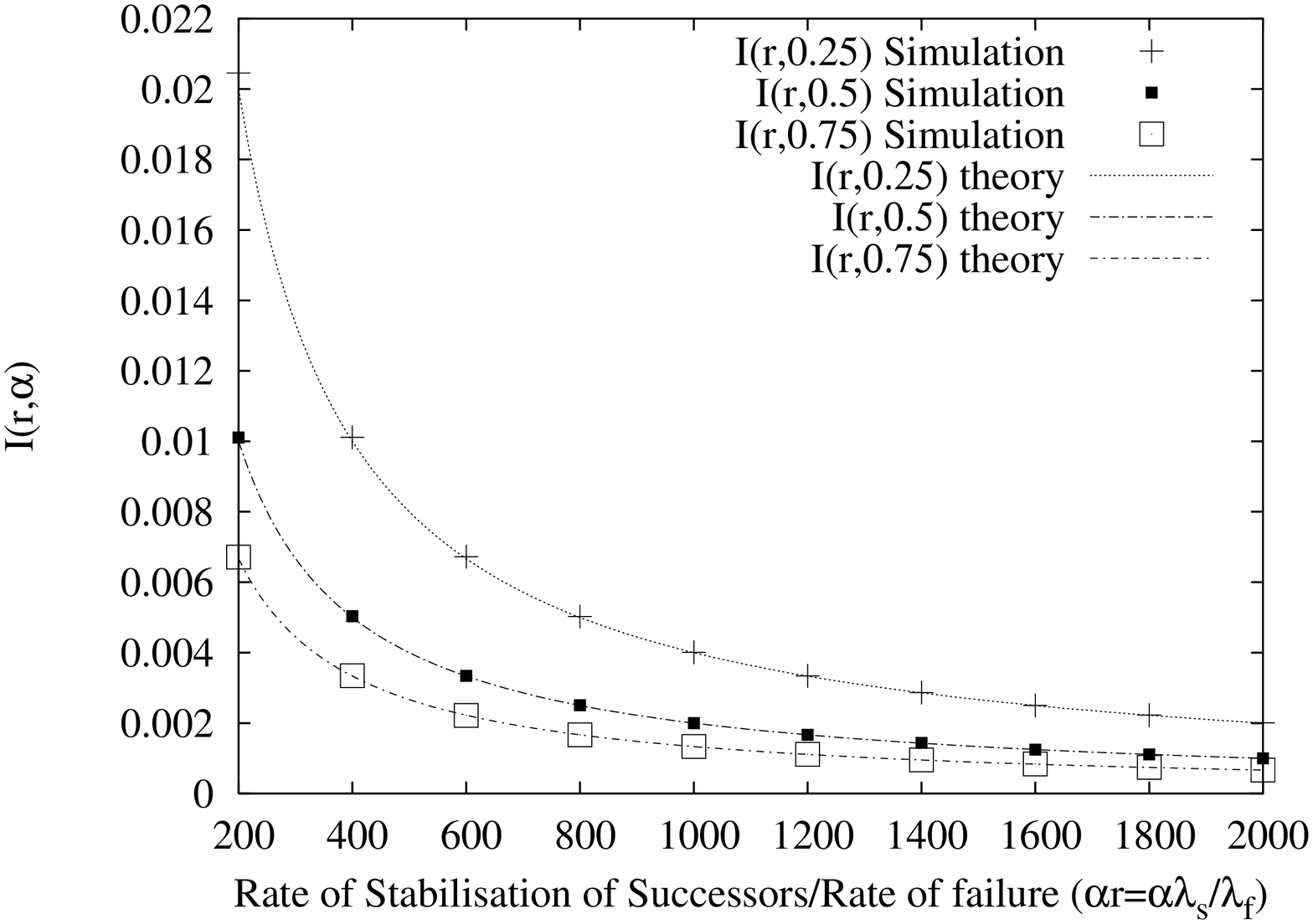}
        \caption{Theory and simulation for inconsistent lookups $I(r,\alpha)$.}
        \label{fig:inconsistent}
\end{figure}

{\bf Lookup Consistency} By the lookup protocol, a lookup is inconsistent if the immediate predecessor of the sought key has a wrong $s_1$ pointer. However, we need only  consider the case when the $s_1$ pointer is pointing to an alive (but incorrect) node since our implementation of the protocol always requires the lookup to return an alive node as an answer to the query. The probability that a lookup is inconsistent $I(r,\alpha)$ is hence $w_1(r,\alpha)- d_1(r,\alpha)$.
This prediction matches the simulation results very well,  as shown in Fig. \ref{fig:inconsistent}. 

\subsection{Failure of Fingers}
\begin{figure}[t]
	\centering
	\includegraphics[width=9cm, height=7cm]{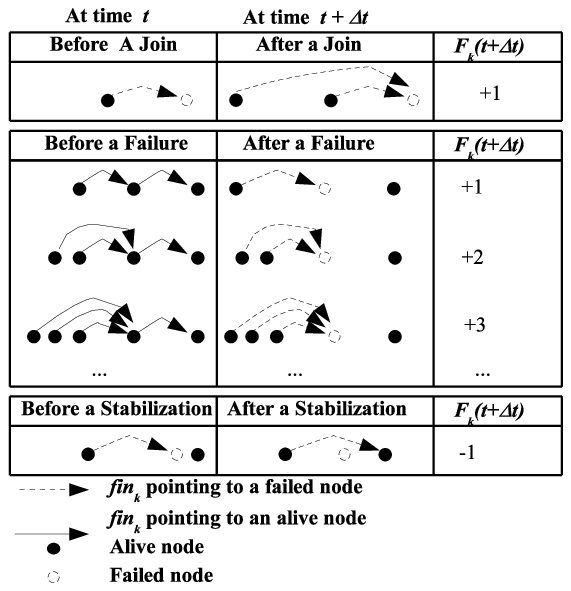}
	\caption{Changes in $F_k$, the number of failed $fin_k$ pointers, due to joins, failures and stabilizations.}
	\label{fig:fk-trans}
\end{figure}

We now turn to estimating the fraction of finger pointers
which point to failed nodes. As we will see this is an important
quantity for predicting lookups, since failed fingers cause timeouts
and increase the lookup length. However, we only need to consider
fingers pointing to {\it dead} nodes.
Unlike members of the successor list,  {\it alive} fingers even if outdated, 
always bring a query closer to the destination and do not 
affect consistency or substantially even the lookup length.
Therefore we consider fingers in only two states, alive 
or dead (failed). By our implementation of the stabilization protocol (see Sections \ref{sec:stab} and \ref{sec:fingers}),
fingers and successors are stabilized entirely
independently of each other to simplify the analysis. 
Thus even though the first finger is
also always the first successor, this information is not used by the node
in updating the finger. Fingers of nodes far apart 
are independent of each other. Fingers of adjacent nodes can be correlated 
and we take this into account. The only assumption in this section is 
in connection with the join protocol as explained below.

Let $f_k(r,\alpha)$ denote the fraction of nodes whose $k^{th}$
finger points to a failed node and $F_k(r,\alpha)$ denote the
respective number. For notational simplicity, we write these as simply
$F_k$ and $f_k$. We can predict this function for any $k$ by again
estimating the gain and loss terms for this quantity, caused by a
join, failure or stabilization event, and keeping only the most
relevant terms. These are listed in Table \ref{tab:f} and illustrated
in Fig. \ref{fig:fk-trans}

\begin{table}
\caption{The relevant gain and loss terms for $F_k$, the number of nodes whose $k{th}$ fingers are pointing to a failed node for $k > 1$.}
\label{tab:f}
	\centering
		\begin{tabular}{|l|l|} \hline
		$F_k(t+\Delta t)$	&  \minorchange{Probability of Occurence}  \\ 
		$= F_k(t)+1$ & $c_{3.1}=(\lambda_j \minorchange{N} \Delta t) \sum_{i=1}^{k}p_{\minorchange{\it join}}(i,k)f_i$
		\\ 
		$= F_k(t)-1$ & $c_{3.2}= (1-\alpha)\frac{1}{{\cal M}}f_k (\lambda_s \minorchange{N} \Delta t)$ \\ 
		$= F_k(t)+1$ & $c_{3.3}= (1-f_k)^2 [1-p_{1}(k)] (\lambda_f \minorchange{N} \Delta t)$ \\ 
		$= F_k(t)+2$ & $c_{3.4}= (1-f_k)^2 (p_{1}(k)-p_{2}(k)) (\lambda_f \minorchange{N} \Delta t)$ \\ 
	  $= F_k(t)+3$ & $c_{3.5}= (1-f_k)^2 (p_{2}(k)-p_{3}(k)) (\lambda_f \minorchange{N} \Delta t)$ \\ 
		$= F_k(t)$   & $1 - (c_{3.1} + c_{3.2} + c_{3.3}+ c_{3.4}+ c_{3.5})$\\ \hline		
		\end{tabular}
\end{table}


A join event can play a role here by increasing the
number of $F_k$ pointers if the successor of the joinee had a failed
$i^{th}$ pointer (occurs with probability $f_i$) and the joinee replicated this from the successor as the joinee's 
\minorchange{$k^{th}$} pointer.
(occurs with probability $p_{\minorchange{\it join}}(i,k)$ 
from property~\ref{prop:copy}). For large enough $k$, 
this probability is one only for $p_{\minorchange{\it join}}(k,k)$, that
is, the new joinee mostly only replicates the successor's $k$th pointer
as its own \minorchange{$k^{th}$} pointer. This is what we consider here.

A stabilization evicts a failed pointer if there was one to begin with.
The stabilization rate is divided by ${\cal M}$, since a node stabilizes
any one finger randomly, every time it decides to stabilize a finger
at rate $(1-\alpha)\lambda_s$.

Given a node $n$ with an alive $k^{th}$ finger (occurs
with probability $1-f_k$), when the node pointed
to by that finger fails, the number of failed $k^{th}$ fingers ($F_k$) increases.
The amount of this increase depends on the number of immediate predecessors of $n$ that were
pointing to the failed node with their $k^{th}$ finger. That number of predecessors could be $0$, $1$, $2$,.. etc.
Using property~\ref{prop:share} the respective probabilities of those cases are: $1-p_{1}(k)$, $p_{1}(k)-p_{2}(k)$, $p_{2}(k)-p_{3}(k)$,... etc.

Solving for $f_k$ in the steady state, we get:
\begin{equation}
\begin{split}
&f_k  =  \frac {
 			\left[2 \tilde{P}_{rep}(k)+2-p_{\minorchange{\it join}}(k)+ \frac{r(1-\alpha)}{{\cal M}} \right]} 
      {2(1+\tilde{P}_{rep}(k))}\\
     &- \frac{
       			\sqrt{ \left[2 \tilde{P}_{rep}(k) + 2 - p_{\minorchange{\it join}}(k)+ \frac{r(1-\alpha)}{{\cal M}} \right]^2 
       - 4(1+\tilde{P}_{rep}(k))^2} 
      }
        {2(1+\tilde{P}_{rep}(k))}
\end{split}
\label{eq:fk}
\end{equation}
where $\tilde{P}_{rep}(k) = \Sigma p_i(k)$.  
In practice, it is enough to keep the first three terms in this sum.
To first order in $\frac{1}{r}$ we have, in analogy to (\ref{eq:wk-leading}),
\begin{equation}
\label{eq:fk-leading}
f_k  \approx  \frac {(1+\tilde{P}_{rep}(k)){\cal M}}{(1-\alpha)r}
\end{equation}
This expression simply says that the fraction of dead fingers
is inversely proportional to the rate of finger stabilizations,
$(1-\alpha)r$, and proportional to how many fingers there are
to stabilize, ${\cal M}$, with the proportionality factor 
$(1+\tilde{P}_{rep}(k))$ depending only on $\rho$. 

To sum up, the computation of the fraction of dead $k^{th}$ finger
pointers is analogous to the calculation of the fraction of
wrong first successor pointer, albeit a bit more involved. No
recursion is involved, in contrast to the calculation of the
fraction of wrong higher successor pointers.
The above expressions, (\ref{eq:fk}) match very well with the simulation  results (Fig. \ref{fig:w}).

\subsection{Cost of Finger Stabilizations and Lookups}
\begin{figure*}[t]
	\centering
	\includegraphics{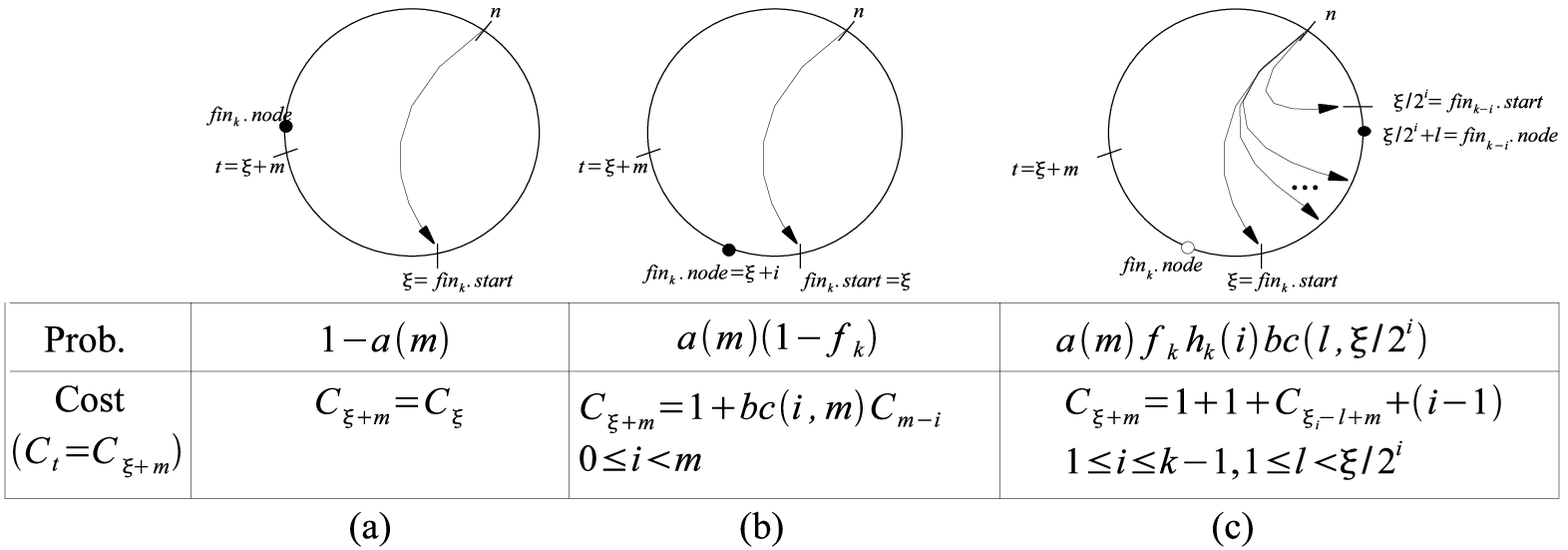}
	\caption{Cases that a lookup can encounter with the respective probabilities and costs.}
	\label{fig:ck}
\end{figure*}
\begin{figure*}[t]
	\centering
		\includegraphics[height=8cm, angle=270]{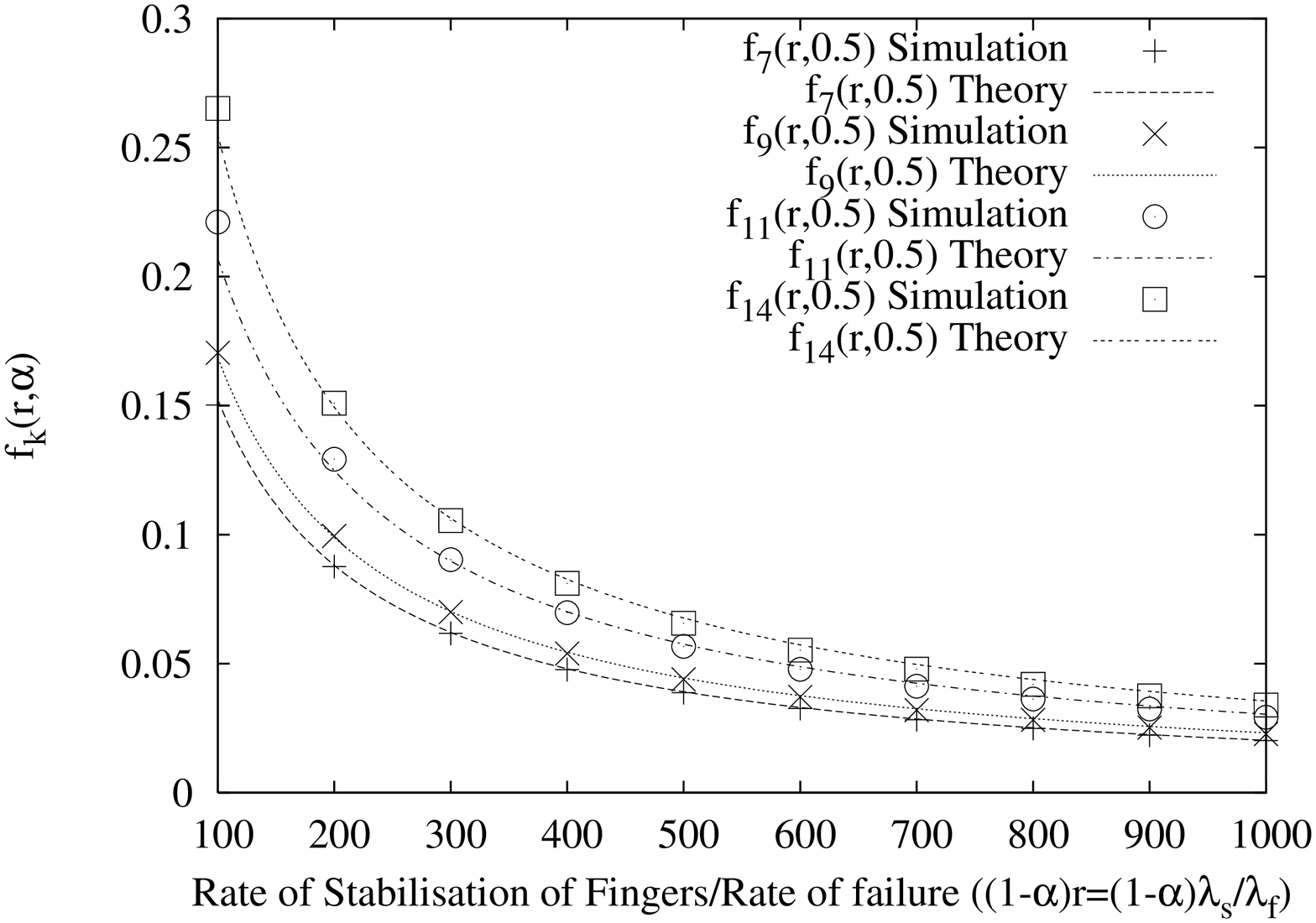}
		\includegraphics[height=8cm, angle=270]{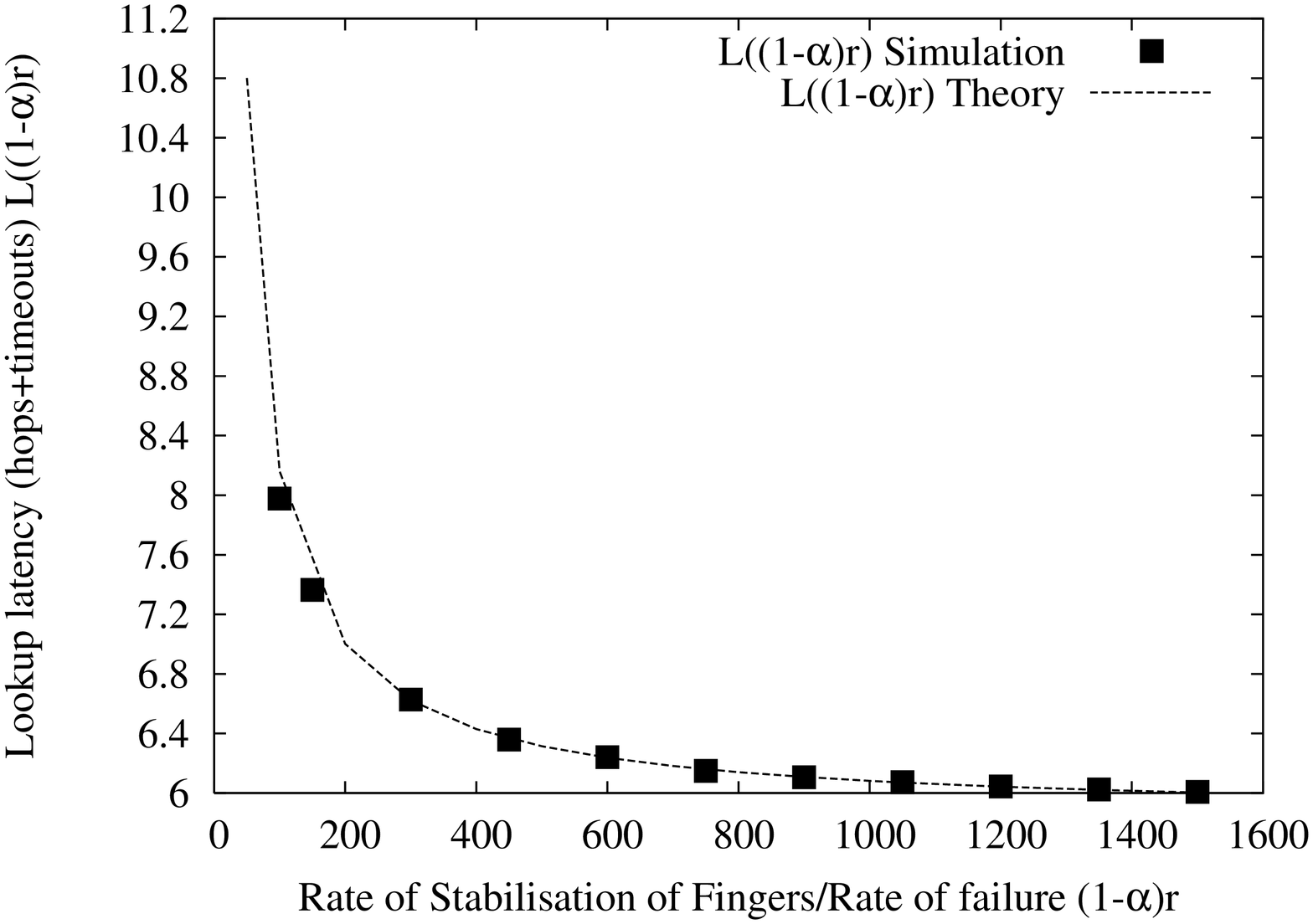}

	\caption{Theory and simulation for probability of failure of the $k^{th}$ finger $f_k(r,\alpha)$, \minorchange{and the lookup length $L(r,\alpha)$.}}
	\label{fig:w}
\end{figure*}

In this section, we demonstrate how the information
about the failed fingers and successors can be used to predict
the cost of stabilizations, lookups or in general the cost for
reaching any key in the id space. By cost we mean the number
of hops needed to reach the destination {\it including }
the number of timeouts encountered en-route. Timeouts occur
every time a query is passed to a dead node. The node does not answer and
the originator of the query has to use another finger instead. 
For this analysis, we consider timeouts and hops to add  equally 
to the cost. We can  easily generalize this analysis to investigate the case 
when a timeout costs some factor $\gamma$ times the cost of a hop. 

Define $C_{t}(r, \alpha)$ (also denoted by $C_{t}$) to be the expected cost for a given node 
to reach some target key which is $t$ keys away from it (which
means reaching the first successor of this key). For example, 
$C_1$ would then be the cost of looking up the adjacent key ($1$
key away). Since the adjacent key is always stored at the
first alive successor, therefore if the first successor is 
alive (which occurs with probability $1-d_1$), the cost will be $1$ hop.
If the first successor is dead but the second is alive (occurs with probability
$d_1(1-d_2)$), the cost will be 1 hop + 1 timeout = $2$ and the \emph{expected} cost is
$2 \times d_1(1-d_2)$ and so forth. Therefore, we have $C_1= 1-d_1 +  2 \times d_1(1-d_2) + 3 \times d_1 d_2 (1-d_3)+ \dots
\approx 1 + d_1 = 1+1/(\alpha r)$. 

To find the expected cost for reaching a general distance $t$ we need
to closely follow the Chord protocol, which would lookup $t$ by first finding 
the closest preceding finger. For the purposes of the analysis,
we will find it easier to think in terms of the closest preceding {\it start}.
Let us hence define $\xi$ to be the {\emph start} of the 
finger (say the $k^{th}$) that most closely precedes $t$. 
Hence $\xi = 2^{k-1} + n$ and
$t = \xi+m$ \textit{i.e.}, there are $m$ keys between the sought target $t$ 
and the start of the closest preceding
finger.  With that, we can write a recursion relation 
for $C_{\xi+m}$ as follows:

\begin{equation}
\label{eq:cost}
\begin{split}
&C_{\xi+m} =  C_{\xi} \left[1-a(m)\right]  						\\
				         &+ (1-f_k) a(m)\left[1 + \sum_{i=0}^{m-1} bc(i,m)C_{m-i}\right]					
          \\					 
					 &+ f_k  a(m) \biggl[ 1 + \sum_{i=1}^{k-1} h_k(i) \\
					 &\sum_{l=0}^{\xi/2^i-1}bc(l,\xi/2^i)(1+(i-1) +C_{\xi_i-l+m}) + O(h_k(k)) \biggr]
\end{split}			
\end{equation}

where $\xi_i \equiv \sum_{m=1,i} \xi/2^{m}$ and $h_k(i)$ is the 
probability that a node is forced to use its $k-i^{th}$ finger owing to the 
death of its $k^{th}$ finger.
The probabilities $a,b,bc$ have already been introduced in Section 
{\ref{sec:internode}},
and we define the probability $h_k(i)$ below.

The lookup equation though rather complicated at first sight 
merely accounts for all the possibilities that
a Chord lookup will encounter, and deals with them 
exactly as the protocol dictates. 

The first term (Fig. \ref{fig:ck}~(a)) accounts for the eventuality that there is no node intervening 
between $\xi$ and $\xi+m$ (occurs with probability $1-a(m)$). 
In this case, the cost of looking for $\xi + m$ is the same
as the cost for looking for $\xi$. 

The second term (Fig. \ref{fig:ck}~(b)) accounts for the situation when a node does intervene in between (with
probability $a(m)$), and this node is alive (with probability $1-f_k$).
Then the query is passed on to this node (with $1$ added to
register the increase in the number of hops) and then the cost depends on
the length of the distance between this node and $t$.

The third term (Fig. \ref{fig:ck}~(c)) accounts for the case when the intervening node is dead
(with probability $f_k$). Then the cost increases by $1$ (for a timeout)
and the query needs to find an alternative 
lower finger that most closely precedes
the target. Let the $k-i^{th}$ finger (for some $i$, $1 \leq i \leq k-1$) 
be such a finger. This happens with probability $h_k(i)$ 
\textit{i.e.}, the probability
that the lookup is passed back to the $k-i^{th}$ finger either because the intervening fingers 
are dead or share the same finger table entry as the $k^{th} $ finger is denoted by $h_k(i)$. 
The start of the $k-i^{th}$ finger is at $\xi/2^i$ and the distance between 
$\xi/2^i$ and $\xi$ is equal to $\sum_{m=1,i} \xi/2^{m}$ 
which we denote by $\xi_i$. 
Therefore, the distance from the {\it start} of the $k-i^{th}$ to the 
target is equal to $\xi_i+m$. 
However, note that $fin_{k-i}.node$ could be $l$
keys away (with probability $bc(l,\xi/2^i)$) from $fin_{k-i}.start$ 
(for some $l$, $0 \leq l < \xi/2^i$).
Therefore, after making one hop 
to $fin_{k-i}.node$,
the remaining distance to the target is  $\xi_i+m-l$.
The increase in cost for this operation is $1+(i-1)$; the $1$ indicates
the cost of taking up the query again by $fin_{k-i}.node$,
and the $i-1$ indicates the cost for trying and discarding each of
the $i-1$ intervening fingers. 
The probability $h_k(i)$ is easy to compute given
property \ref{prop:ab} and the expression
for the $f_k$'s computed in the previous section.

\begin{equation}
\label{eq:hki}
\begin{split}
h_k(i) = & a(\xi/2^{i}) (1-f_{k-i})  \\
       \times &\Pi_{s=1,i-1} (1-a(\xi/2^{s}) + a(\xi/2^s)f_{k-s}), i<k \\
h_k(k) = & \Pi_{s=1,k-1} (1-a(\xi/2^{s}) + a(\xi/2^s)f_{k-s})  
\end{split}
\end{equation}

In (\ref{eq:hki}) we account for all the
reasons that a  node may have to use its $k-i^{th}$ finger 
instead of its $k^{th}$ finger. This could happen because the 
intervening fingers were either dead or not distinct. 
The probabilities $h_k(i)$ satisfy the constraint $\sum_{i=1}^{k} h_k(i)=1$
since clearly, either a node uses any one of its fingers
or it doesn't. This latter probability is $h_k(k)$, that is the probability that a node
cannot use any earlier entry in its finger table. 
In this case, $n$ proceeds to its successor list. 
The query is now passed on to the first alive successor 
and the new cost is a function of the distance of this node 
from the target $t$.
We indicate this case by the last term in \ref{eq:cost} which is 
$O(h_k(k))$. This can again be computed from the inter-node distribution
and from the functions $d_k(r,\alpha)$ computed earlier.
However in practice, the probability for this
is extremely small except for targets very close to $n$. 
Hence this does not
significantly affect the value of general lookups and we ignore it
in our analysis.

%
%
%

The cost for general lookups is hence 
$$
L(r,\alpha) = \frac{\Sigma_{i=1}^{{\cal K} -1} C_i(r,\alpha)}{\cal K} 
$$

The lookup equation is solved recursively numerically, 
given the coefficients
and $C_1$. 
In Fig.~\ref{fig:w}, we compare  theoretical results with
simulation for $N=1000$. It is seen that the theory matches 
the simulation results very well.

\begin{figure}[t]
	\centering
	\includegraphics[height=8cm, angle=270]{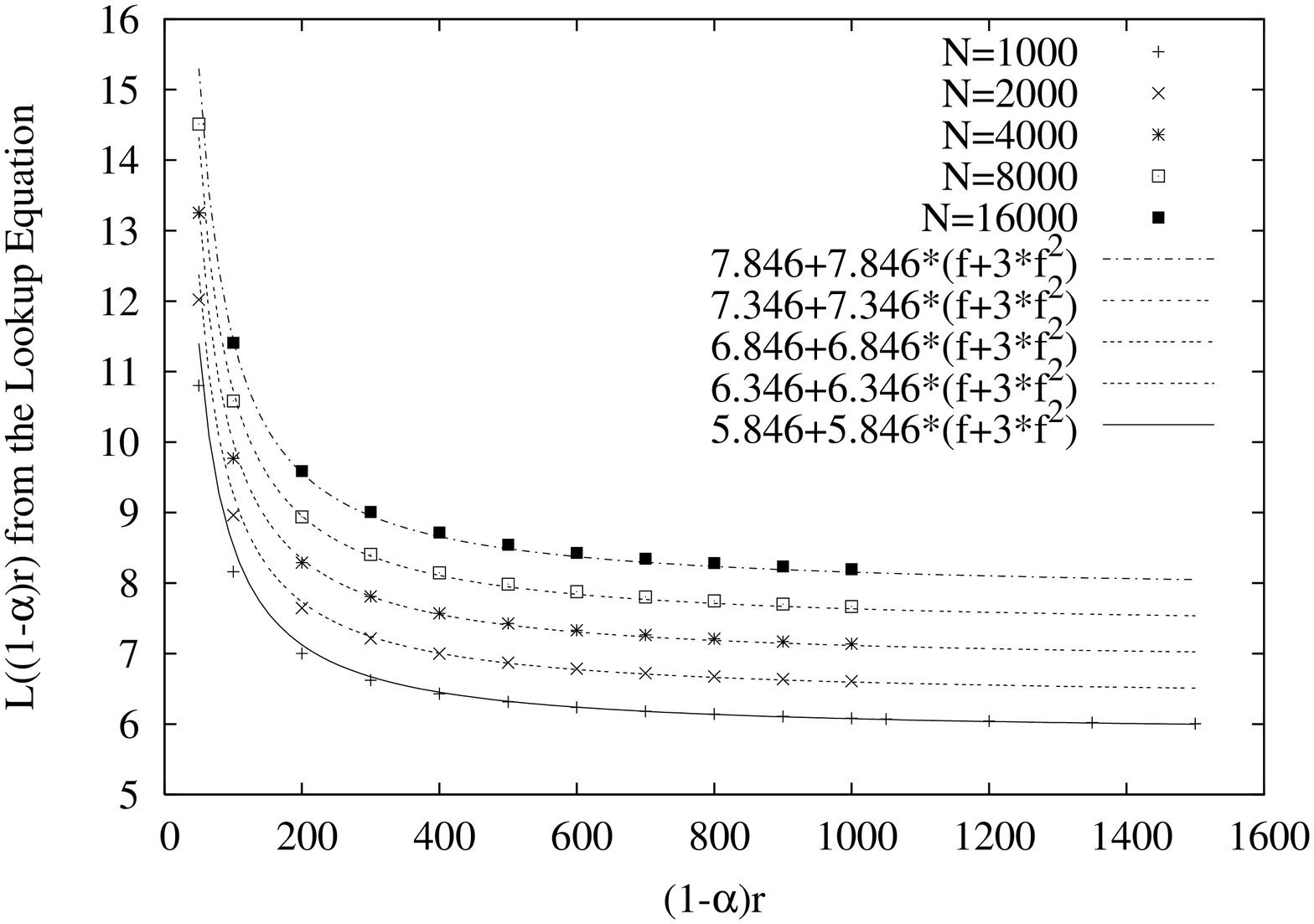}
	\caption{Lookup cost, theoretical curve, for \minorchange{$1000$,$2000$,$4000$,$8000$ and $16000$} peers. 
The rationale for the fits is explained in the text.}
	\label{fig:lookup_theory}
\end{figure}

In Fig.~\ref{fig:lookup_theory} we also show the theoretical predictions
for some larger values of $N$. \minorchange{From the structure of 
Equation \ref{eq:cost}, it is clear that the dependence of the average lookup on churn comes entirely 
from the presence of the terms $f_k$. Since $f_k \sim f$ is independent of $k$ for large fingers, we can
approximate the average lookup length by the  
functional form $L (r, \alpha) = A + {B}f + C f^{2} + \cdots $.
The coefficients $A, B, C$ {\it etc} can be recursively computed by solving the lookup equation
to the required order in $f$ and depend only on $N$ the number of nodes, $1- \rho $ the  
density of peers and $b$ the base or equivalently the 
size of the finger table of each node. 
The advantage of writing the lookup length this 
way is that churn-specific details such as how new joinees construct a finger table or how exactly
stabilizations are done in the system, can be isolated in the expression for $f$.
If we were to change our stabilization strategy for example \cite{KEAH_inprep}, we could immediately
estimate the lookup length by plugging in the new expression for $f$ in the above relation.}

\minorchange{The coefficient $A$, which is the lookup cost without churn 
can be obtained very precisely for any base $b$}, from analyzing
(\ref{eq:cost}) in the zero-churn case. This analysis is rather 
laborious and will be presented elsewhere \minorchange{\cite{KEAH_inprep}}. 
It confirms the well-known result
$A= \frac{1}{2}\log_2 N $ and in addition
reproduces small deviations from this behavior previously
observed by us in numerical simulations \cite{elAnsaryAurellHaridi}.
The values of $A$ in Fig.~\ref{fig:lookup_theory} are
taken from this analysis.

$B$ can be qualitatively estimated as follows :
every sufficiently long finger is dead with some
finite probability \minorchange{$f$ given by (\ref{eq:fk}).
If $A$ is the average value of the lookup length {\it without} churn, then
each look-up encounters $f A$ dead fingers on average}. This estimate
predicts a look-up cost of approximately \minorchange{$A (1+f)$, giving $B=A$ and $C$ and all other 
coefficients equal to $0$.}.

\minorchange {In Fig.~\ref{fig:lookup_theory} we show that the best fit to the data is obtained in fact
by taking $B=A$ and $C=3A$. The expression for $f$ is taken from \ref{eq:fk} for large $k$ (for a system with $20$ fingers, 
the expression for $f_k$ becomes independent of $k$ for $k \ge 13$). 
In general, as mentioned earlier, $B$ and $C$ can be obtained accurately for any value of the system parameters
by the numerical solution of Eq. ~\ref{eq:cost} to the required order}.


\section{Discussion and Conclusion}
In this paper  we have presented a detailed theoretical
analysis of a DHT-based P2P system, Chord, using a fluid model.
The technique for deriving the fluid model has been borrowed
from the master equation approach of physics, which helps in
systematically taking different dynamical effects into account.
This analysis differs from previous theoretical
work done on DHTs in that it aims not at establishing bounds, but on
precise determination of the relevant quantities in this 
dynamically evolving system. From the match of our theory and
the simulations, it can be seen that we can predict with an accuracy
of greater than $1\%$ in most cases.
Though this analysis is not {\it exact}, since it 
takes only some (and not all) correlations into account, yet
it provides a methodology for keeping track of most of the relevant
details of the system. We expect that a \minorchange{similar}
analysis can be done for most other DHT's, thus
helping to establish quantitative guidelines for their
comparison.

The main conclusions for the analysis of Chord in a statistically
steady state are the following.

\begin{property}
\label{prop:one-over-r}
As a function of $r$, the ratio of the rate of stabilizations
to the rate of failures, the fraction of wrong pointers of
any kind (successors or fingers) 
is to leading order and good approximation 
$\hbox{Const.}/r$, where the constant depends on the pointer.   
\end{property}

\begin{property}
\label{prop:break-up}
The probability of break up of a ring can be estimated
from the knowledge of the fraction of wrong first
successors, wrong second successors, etc. This probability is
generally very low when every node has a sufficient number of successors, 
indicating that Chord is robust against ring break-up.
\end{property}

\begin{property}
\label{prop:dep-on-k}
At a given value of $r$, the fraction of wrong successors, $w_k$, and
the fraction of dead fingers, $f_k$, increases with $k$. 
The fraction of wrong successors increases indefinitely, and becomes
of order one at $k$ about $\sqrt{r}$ \minorchange{for the particular stabilization strategy that we have used.} 
The fraction of dead fingers on the other hand tends to a constant for sufficiently large $k$.
\end{property}

\begin{property}
\label{prop:cost}
The look-up cost, which is the expected number of hops including time-outs,
can be computed by numerical recursion. The fraction of incorrect 
finger pointers $f_k$ \minorchange{($\sim f$ for large $k$)} 
is a required input for this recursion. 
The lookup cost tends to the
well-known average number of hops without churn when 
\minorchange{$f$ is small (or churn is low)
and increases when $f$ is large. We show that it can be well described by the formula $A(1+ g(f))$, where 
$A$ is the value of the lookup cost without churn and $g(f)$ is well approximated by $f + 3f^{2}$ for 
$N << K$. In general $g(f)$ can be obtained accurately to any desired order by solving Eq. \ref{eq:cost} recursively
to the required order in $f$.}
\end{property}

\begin{property}
\label{prop:scale}
The preceding note brings out the following
simple feature of Chord: under any state of churn,
sufficiently long fingers are all dead with essentially the
same probability. Hence, in a sufficiently large system,
a look-up will almost surely encounter one or more
dead fingers, leading to time-outs.
For applications where time-outs should be the
exception and not the norm, this
paper helps in estimating how much stabilization is necessary
under a given level of churn, to achieve such a level of performance.
\end{property}

\minorchange{
\begin{property}
\label{prop:general}
The preceding note also brings out the additional feature
that by writing the lookup cost in the above simplified form, we can isolate the effects of 
churn-specific details in the expression for $f$. Changing details in the 
join protocol or changing the maintenance strategy \cite{KEAH_inprep} merely cause a change in the expression 
for $f$. The lookup cost with this new strategy can then be immediately assessed  for any $r$, by plugging in the new expression for $f$ in the expression for the lookup cost 
(as opposed to solving Eq. \ref{eq:cost} each time for each value of $r$).
\end{property}
}

The impact of this work can be summarized as follows: given that periodic 
stabilization is a fundamental technique for topology maintenance 
in DHTs, the question: "How often should 
a DHT node perform periodic stabilization?" is of great practical relevance. 
The answer to this question depends on several factors.
First we need to know where the DHT is deployed, in a LAN, 
in a cooperative milieu,  or among public non-trusting partners, 
\textit{i.e.}, what is the expected join/failure rate (churn)?  
Secondly, since DHTs involve different types of stabilizations, we need to
know which of these rates is of interest to optimize. For example,
in the  DHT studied in this paper,  there is both ring stabilization as well 
as finger stabilization.  Thirdly, we also need to know 
whether  we have performance goals which require us to know
how much stabilization is needed, or constraints on bandwidth
which necessitate a knowledge of the expected performance.
Previous analytical attempts (see Section \ref{sec:related}) 
have addressed these question through 
the identification of general (algorithm/system-neutral) 
bounds on stabilization rates. 

In this paper, we have taken another point of view. We 
have traded-off generality for accuracy.
That is, we have produced results that can describe to a 
very high degree of accuracy  
quantities like the probability of inconsistent look-ups and the
 expected look-up length as functions of the stabilization and churn rates. 
\minorchange{Many of the insights we get from this analysis
such as most of the points listed above, would be very hard to
come by from simulations alone.}
So for instance, the  formulae produced in this paper could 
directly be used by a 
system administrator  or the person in charge of deploying a DHT 
as a guide for configuring stabilization rates. 
While the results are based on Chord, all analyses 
concerning the ring (break-up and 
inconsistency) are applicable to many other systems, since 
consistent hashing on a ring is a 
recurring component in many other DHTs. 

\minorchange{
\section{Limitations and Future Work}
The main limitation of this work stems from the fact that the results are
inherently dependent on the intricate details of the analyzed algorithms. While some changes in 
the algorithms can be easily accommodated without redoing the analysis (as 
explained in \ref{prop:general}), others such as a different lookup strategy or a different placement 
of fingers would necessitate recalculating all the quantities again. However, results concerning
the ring-related aspects like successor lists, break-up probability and inter-node distributions are 
likely to be reusable in other variations of the Chord protocols as
well other systems using a ring geometry.

For the future, the authors' research agenda include the introduction 
of extensions to the current model to be able to account for locality-awareness
and different topology maintenance techniques. Some work towards the latter goal has already been done in 
\cite{KEAH_inprep}. Relatedly, a useful application for this work is to enable systems to dynamically
self-tune their stabilization rates and  choose the best maintenance technique to achieve a desired hop count.  
}

{\footnotesize
\bibliographystyle{amsplain}
\bibliography{P2P_v2}

\providecommand{\bysame}{\leavevmode\hbox to3em{\hrulefill}\thinspace}
\providecommand{\MR}{\relax\ifhmode\unskip\space\fi MR }
\providecommand{\MRhref}[2]{%
  \href{http://www.ams.org/mathscinet-getitem?mr=#1}{#2}
}
\providecommand{\href}[2]{#2}
\begin{thebibliography}{10}

\bibitem{aberer04tok}
Karl Aberer, Anwitaman Datta, and Manfred Hauswirth, \emph{Efficient,
  self-contained handling of identity in peer-to-peer systems}, IEEE
  Transactions on Knowledge and Data Engineering \textbf{16} (2004), no.~7,
  858--869.

\bibitem{AnickMitraSondhi}
D.~Anick, D.~Mitra, and M.M. Sondhi, \emph{Stochastic theory of data-handling
  systems with multiple sources}, Bell Systems Technical Journal \textbf{61}
  (1982), 1871--1894.

\bibitem{aspnes02FaultTolerant}
James Aspnes, Zo{\"e} Diamadi, and Gauri Shah, \emph{Fault-tolerant routing in
  peer-to-peer systems}, Proceedings of the twenty-first annual symposium on
  Principles of distributed computing, ACM Press, 2002, pp.~223--232.

\bibitem{Erlang}
E.~Brockmeyer, H.L. Halstrom, and Arns Jensen, \emph{{The life and works of
  A.K. Erlang}}, The Copenhagen Telephone Company, 1948.

\bibitem{rowstron04depend}
Miguel Castro, Manuel Costa, and Antony Rowstron, \emph{Performance and
  dependability of structured peer-to-peer overlays}, Proceedings of the 2004
  International Conference on Dependable Systems and Networks (DSN'04), IEEE
  Computer Society, 2004.

\bibitem{clevenot}
Florence Cl\'evenot and Philippe Nain, \emph{A simple fluid model for the
  analysis of the squirrel peer-to-peer caching system}, IEEE INFOCOM 2004,
  2004.

\bibitem{elAnsaryAurellHaridi}
Sameh El-Ansary, Erik Aurell, and Seif Haridi, \emph{A physics-inspired
  performace evaluation of a structured peer-to-peer overlay network}, {The
  International Conference on Parallel and Distributed Computing and Networks
  (PDCN 2005)}, 2005.

\bibitem{KEAH1}
Supriya Krishnamurthy, Sameh El-Ansary, Erik Aurell, and Seif Haridi, \emph{A
  statistical theory of chord under churn}, The 4th International Workshop on
  Peer-to-Peer Systems (IPTPS'05) (Ithaca, New York), February 2005.

\bibitem{KEAH_inprep}
\bysame, \emph{Comparing maintenance strategies for overlays}, Tech. report,
  Swedish Institute of Computer Science, in preparation 2007.

\bibitem{dhtcomparison:infocom05}
Jinyang Li, Jeremy Stribling, Robert Morris, M.~Frans Kaashoek, and Thomer~M.
  Gil, \emph{A performance vs. cost framework for evaluating dht design
  tradeoffs under churn}, Proceedings of the 24th Infocom (Miami, FL), March
  2005.

\bibitem{nowell02analysis}
David Liben-Nowell, Hari Balakrishnan, and David Karger, \emph{Analysis of the
  evolution of peer-to-peer systems}, ACM Conf. on Principles of Distributed
  Computing (PODC) (Monterey, CA), July 2002.

\bibitem{vanKampen}
{N.G. van Kampen}, \emph{{Stochastic Processes in Physics and Chemistry}},
  {North-Holland Publishing Company}, 1981, ISBN-0-444-86200-5.

\bibitem{qiu}
Dongyu Qui and R.~Srikant, \emph{Modeling and performance analysis of
  bittorrent-like peer-to-peer networks}, SIGCOMM'04 (Portland, Oregon), August
  2004.

\bibitem{rhea04handling}
Sean Rhea, Dennis Geels, Timothy Roscoe, and John Kubiatowicz, \emph{Handling
  churn in a {DHT}}, Proceedings of the 2004 USENIX Annual Technical
  Conference(USENIX '04) (Boston, Massachusetts, USA), June 2004.

\bibitem{chord:ton}
Ion Stoica, Robert Morris, David Liben-Nowell, David Karger, M.~Frans Kaashoek,
  Frank Dabek, and Hari Balakrishnan, \emph{Chord: A scalable peer-to-peer
  lookup service for internet applications}, {IEEE} Transactions on Networking
  \textbf{11} (2003).

\bibitem{wang03resilience}
Shengquan Wang, Dong Xuan, and Wei Zhao, \emph{On resilience of structured
  peer-to-peer systems}, GLOBECOM 2003 - IEEE Global Telecommunications
  Conference, Dec 2003, pp.~3851--3856.

\end{thebibliography}
}

\end{document}